\documentclass[12pt,letterpaper]{article}
\usepackage{graphicx}
\usepackage{epstopdf}
\usepackage{amsmath}
\usepackage{amsfonts}
\usepackage{amssymb}

\usepackage{graphicx}

\def\ee{\end{eqnarray}}

\def\d{\partial}

\newcommand{\be}{\begin{eqnarray}}
\newcommand{\en}{\end{eqnarray}}
\newcommand{\bea}[1]{\left(\begin{array}{#1}}
\newcommand{\ena}{\end{array}\right)}

\begin{document}

%\begin{flushright} {\footnotesize IC/2007/001}  \end{flushright}
\vspace{5mm}
\vspace{0.5cm}
\begin{center}

\def\thefootnote{\fnsymbol{footnote}}

{\Large \bf Cosmological Perturbations at Second Order \\ 
and \\[0.30cm]
Recombination Perturbed}
\\[0.5cm]
{\large  Leonardo Senatore$^{a,b,c}$, Svetlin Tassev$^c$ and Matias Zaldarriaga$^{b,c}$}
\\[0.5cm]

\vspace{.3cm}
{\normalsize { \sl $^{a}$ School of Natural Sciences, Institute for Advanced Study, \\Olden Lane, Princeton, NJ 08540, USA}}\\

\vspace{.3cm}
{\normalsize { \sl $^{b}$ Jefferson Physical Laboratory, Harvard University, Cambridge, MA 02138, USA}}\\

\vspace{.3cm}
{\normalsize {\sl $^{\rm c}$  
Center for Astrophysics, Harvard University, Cambridge, MA 02138, USA}}

%\vspace{.2cm}

%{\small \textit{$^{\rm b}$ Abdus Salam International Center for
%Theoretical Physics\\ Strada Costiera 11, 34014 Trieste, Italy}}

\end{center}

\vspace{.8cm}

\hrule \vspace{0.3cm}
{\small  \noindent \textbf{Abstract} \\[0.3cm]
\noindent
We derive the full set of second-order equations governing the evolution
of cosmological perturbations, including the effects of the
first-order electron number density perturbations, $\delta_e$. We
provide a detailed analysis of the perturbations to the recombination
history of the universe and show that a perturbed version of the
Peebles effective 3-level atom is sufficient for obtaining the
evolution of $\delta_e$ for comoving wavenumbers smaller than 1 Mpc$^{-1}$. We calculate rigorously the perturbations to
the Ly$\alpha$ escape probability and show that to a good
approximation it is governed by the local baryon velocity divergence.
For modes shorter than the photon diffusion scale, we find that
$\delta_e$ is enhanced during recombination by a factor of roughly 5
relative to other first-order quantities sourcing the CMB anisotropies
at second order. Using these results, in a companion paper we calculate the CMB bispectrum generated during recombination.
\vspace{0.5cm}  \hrule
\def\thefootnote{\arabic{footnote}}
\setcounter{footnote}{0}

%\documentclass[12pt,letterpaper]{iopart}

%\usepackage{amsmath}

%\usepackage{amsfonts}
%\usepackage{amssymb}
%\usepackage{graphicx}
%\usepackage{bm}

%\setlength{\textwidth}{425pt}
%\setlength{\textwidth}{475pt}
%\setlength{\textheight}{595pt}
%\setlength{\topmargin}{-1.2cm}
%\setlength{\textheight}{655pt}
%\setlength{\oddsidemargin}{-14pt}
%\linespread{1.1}

%\begin{document}

%\begin{flushright} {\footnotesize IC/2007/001}  \end{flushright}

%\vspace{.8cm}

%\hrule \vspace{0.3cm}
%{\small  \noindent \textbf{Abstract} \\[0.3cm]
%\noindent
%Abstract
%\vspace{0.5cm}  \hrule
%\renewcommand{\thefootnote}{\arabic{footnote}}
%\renewcommand{\thefootnote}{\arabic{footnote}}
%\setcounter{footnote}{1}

%\documentclass[letterpaper,aps,secnumarabic,nobalancelastpage,amsmath,amssymb]{revtex4}
%%\documentclass[12pt,preprint,letterpaper,aps,secnumarabic,nobalancelastpage,amsmath,amssymb]{aastex}
%\usepackage{graphics}
%\usepackage{graphicx}
%\usepackage{longtable}
%\usepackage{url}
%\usepackage{bm}
%\begin{document}
\def\be{\begin{eqnarray}}
\def\ee{\end{eqnarray}}
\def\bee{\begin{equation}}
\def\eee{\end{equation}}

%\title{Perturbing Recombination and Cosmological Perturbations at Second Order}
%\author         {Svetlin V. Tassev}
%\email          {stassev@cfa.harvard.edu} 
%%\affil {Harvard-Smithsonian Center for Astrophysics}
%\affiliation {Harvard-Smithsonian Center for Astrophysics

\section{Introduction}
In the last few years there has been great progress in
understanding the non-Gaussianity of the primordial spectrum
of density fluctuations. Starting from the first full
computation of the non-Gaussian features in single field slow roll
inflation \cite{Maldacena:2002vr,Acquaviva:2002ud}, several alternative models
have been proposed that produce a large and in principle detectable
level of non-Gaussianities through different mechanisms for generating density
fluctuations in the quasi de Sitter inflationary phase, both in the case of single field inflation
\cite{Arkani-Hamed:2003uz,Alishahiha:2004eh,Shandera:2006ax,Chen:2006nt,Cheung:2007st}, and in the case of multi-field inflation \cite{Lyth:2002my,Zaldarriaga:2003my}. Further, it has been at last developed a consistent model of bouncing universe \cite{Lehners:2007ac,Buchbinder:2007ad,Creminelli:2007aq} which, though clearly less compelling than inflation, predicts a possibly detectable amount of non-Gaussianity \cite{Creminelli:2007aq}. 
 If large non-Gaussianities will be detected, we would have to abandon the standard slow-roll inflation picture, but would be left with new information to understand the dynamics of the inflaton \cite{Babich:2004gb,Cheung:2007st}.

At the same
time, from the experimental side, the WMAP satellite has allowed
for a huge improvement in our measurement of the properties
of the CMB. Observations seem to confirm the generic predictions
of standard slow roll inflation \cite{Komatsu:2008hk}. Limits on the
primordial non-Gaussianity of the CMB have been significantly improved
\cite{Creminelli:2005hu,Komatsu:2008hk},
but for the moment the data are consistent with a non-Gaussian signal.
Recently it has been realized that even large scale structure surveys are relevantly sensitive to at least some particular kind of non-Gaussianities \cite{Dalal:2007cu,Slosar:2008hx}, giving constraints comparable to those from WMAP \cite{Slosar:2008hx}.

Oversimplifying, the limits on non-Gaussianities are set on the parameter $f_{\rm NL}$ such that the curvature $\zeta$ that we observe in the CMB is given by:
\bee\label{eq:non-gauss-naive}
\zeta(\vec x)=\zeta_g(\vec x)-\frac{3}{5}f_{\rm NL}\left( \zeta_g(\vec x)^2-\langle\zeta_g^2\rangle\right)\ ,
\eee
where $\zeta_g(\vec x)$ is a gaussian random variable~\footnote{This oversimplifies the structure of the possible non-Gaussianities \cite{Babich:2004gb,Cheung:2007st}, but this is enough for our introductory purposes.}. So far we have a 1-$\sigma$ error on $f_{\rm NL}$ of order $\Delta f_{\rm NL}\simeq 30$ from WMAP \cite{Komatsu:2008hk}, and soon the Planck satellite is expected to improve this limit to $\Delta f_{\rm NL}\lesssim 4$ \cite{Babich:2004yc}. However, it is clear by looking at eq.~(\ref{eq:non-gauss-naive}) that the corrections due to the non-linearities of general relativity and of the Boltzmann equations that regulate the fluid dynamics are very naively expected to give rise to $f_{\rm NL}$ of ${\cal{O}}(1)$. This is clearly very close to what Planck will achieve, and it means that in order to be able to fully exploit the next new experimental results, the calculation of the non-Gaussianity induced by the non-linearities in general relativity and in the plasma physics needs to be done.

Of course we are not the first ones to realize the importance of this issue. In particular, an effort of finding and solving the full set of second order equations is on-going by some groups \cite{bartolo, bartolo2, astro-ph/0703496,Pitrou:2008ak}. For example, the three point function of temperature fluctuations in the particular limit where one of the wavelengths is much longer than the horizon at recomination and than the other two wavelengths has been consistently computed, finding an effect of order $f_{\rm NL}\sim1$~\cite{Creminelli:2004pv}. This is still far from the required full calculation. The derivation of the second order equation requires to deal with some subtleties such as the formulation of the Boltzmann equation in curved spacetime that were neglected in former derivations such as the one of \cite{bartolo}, and that could affect the result at the order we care about. One of the two purposes of the present paper is to derive in a consistent way this set of equations. Unfortunately this will take us through a rather technical path, though we will try to make always clear the physical interpretation of what we are doing. We can already synthesize the main points of the derivation. We will work in the all-orders generalization of the Newtonian gauge, which is called Poisson Gauge. Writing Einstein equations in this gauge is tedious, but not particularly difficult. Writing down the Boltzmann equations in this frame is instead a little more subtle. The Boltzmann equation consists of a free-streeming term (so called Liouville term) which takes into account of the free evolution of the species one-particle distribution, and of a collision term, which takes into account how interactions among the particles affect the distribution. The collision term is expressed in terms of cross-sections that are measured and well expressible in Minkowski space. This means that, in order to write down the Boltzmann equation in Poisson gauge, one needs at each point  to find the coordinate transformation to the local inertial frame, write down the Boltzmann equation there, and then transform back to the original frame. This will give us the recipe to write down the Boltzmann equations at second order in the perturbations. 

If schematically for the moment we write each perturbation $\delta$ as $\delta=\delta^{(1)}+\delta^{(2)}$ where $\delta^{(2)}$ is of order $(\delta^{(1)}){}^2$, the resulting first and second order set of Einstein and Boltzmann equations will take the form of a system of equations of the form:
\be
&&D[\delta^{(1)}]=0\ , \\ \nonumber
&&D[\delta^{(2)}]=S[\delta^{(1)}{}^2]\ .
\ee
Here $D$ is a differential operator that is the same both for the first and the second order equations. The second order equation, unlike the first order one, has a source term $S$ proportional to the square of the first order perturbations \footnote{The first order perturbations are non zero only once one assigns them some primordial non-zero initial conditions.}. 

Among the many terms that contribute to the source $S$, there are some (coming from the Compton scattering collision term) in which one of the two first order perturbations is the perturbation in the number density $n_e$ of free electrons: $\delta_e\equiv \delta n_e/n_e$. If one concentrates, as we will do, on the CMB, then $\delta_e$ affects the CMB anisotropy only at second order. This can be seen in the following way. The free electron density affects the CMB temperature through the Compton collision term, which determines the visibility function and the diffusion scale. The former gives the probability for a photon to originate from a given distance along the line of sight to the observer, while the latter accounts for the Silk damping at higher multipoles $l$'s. In a homogeneous universe, before, during and after recombination the radiation temperature decreases as $a^{-1}$, $a$ being the scale factor, irrespective of the electron density. Thus, a perturbation to the electron density changes the position of the last scaterring surface and the mean free path of the photons before decoupling, but not the observed radiation temperature. Therefore, since at first order we would perturb $n_e$ and keep the other quantities unperturbed, the CMB anisotropies are not affected by $\delta_e$ at this order. However, they clearly are so at second order.
This explains why this quantity had never been computed before in detail. The first part of the paper will in fact be devoted to the computation of this quantity which is necessary to solve for the second order perturbations. As we will see, this task will not be completely straightforward because of the way recombination occurs in our universe.  In particular, even in the homogeneous universe the collision term which controls the population of free electrons does not vanish (of course, otherwise there would be no recombination in the homogeneous universe). This means that the same subtleties that we mentioned above about writing down the Boltzmann equation at second order will now apply also at first order. Further, even in the unperturbed case, recombination is not a very straightforward process, involving atomic transition between many different states. The treatment is simplified by the fact that matter and radiation are very close to equilibrium during recombination, with the most notable exception being the ground and first excited states of Hydrogen which are not mutually in equilibrium. This results in a series of approximations, which allow one to treat Hydrogen as an effective 3-level atom (ground state, first excited state and continuum) as was done in the classic work of Peebles \cite{peebles}. In the end, one can write a single ordinary differential equation (ODE) governing the evolution of the electron density. Working in the same approximations, we shall derive the analogous equation for the perturbation of the electron density~eq.~(\ref{eq:delta_e_first}). We also derive the equation for the perturbations of the matter temperature~eq.~(\ref{deltaTM1})  \footnote{One can try to directly perturb Peebles' ODE and write an equation for $\delta_e$ which then has to be complemented with an equation for the perturbations to the matter temperature, $\delta_{T_M}$. The calculation of $\delta_e$ has been already attempted in \cite{Novosyadlyj} without accounting for the effects of metric perturbations and of the perturbation to the escape probability, and in \cite{arXiv:0707.2727v1,astro-ph/0702600v2}, where only a non rigorous  derivation of the perturbations to escape probability was given. We will show that the formulas given in \cite{arXiv:0707.2727v1,astro-ph/0702600v2} are indeed correct apart for irrelevant corrections.}. 
%I needs to be complemented with an equation for the perturbations of the matter and radiation temperatures~eq.~(\ref{}) and (\ref{}) {\bf (??????????????)} .

At least on large scales, one can consider the primordial perturbation as representing a time delay $\delta t$ of the unperturbed solution in each point in space (of the order of $10^{-5} H^{-1}$). In this case, $\delta n_e\sim \dot n_e \delta t$, and since the time-scale of recombination is rather quick, $\dot n_e/n_e\sim 15\, H$ (that is 5 times the expansion rate), we expect $\delta_e\sim 5 \times 10^{-5}$, an enhanced perturbation with respect to the standard ones. This is in fact what will be the result of a detailed calculation.

Because of this enhancement, it is natural to wonder if this might affect the CMB, for example by generating an $f_{\rm NL}$ of order 5, which could be potentially detectable by Planck.  We find the way $\delta_e$ changes the CMB is mainly by delaying the time of recombination, by modifying the shape of the visibility function, and by changing the diffusion scale. The result is that $\delta_e$  gives a potentially detectable effect equivalent to $f_{\rm NL}\sim 5$. Since the calculation and the interpretation of this result is not completely straightforward, it will be  the subject of a companion paper \cite{inprep}.

The paper is organized as follows. In sec.~\ref{section:review} we review recombination in the unperturbed universe. In sec.~\ref{section:firstorder} we derive the first order equations for recombination in the perturbed universe and discuss the results of their numerical integration. In sec.~\ref{section:secondorder} we derive the full set of second order equations for cosmological perturbations. In sec.~\ref{section:summary} we summarize our results. In App.~\ref{section:A} we review the formulation of kinetic theory in curved spacetime. In App.~\ref{section:B} we derive in a rigorous way the perturbations to the escape probability. In App.~\ref{section:C} we give an erratum to the Compton collision term at second order which is already present in the literature.

\section{Review of standard recombination}\label{section:review}
Here we review the standard Peebles' \cite{peebles} treatment of recombination in the homogeneous universe. We will also state what are the approximations that are implicitly done in this treatment. When dealing with cosmological perturbations, there will be additional timescales corresponding to the frequency of the various modes, and we will have to check in which regime the standard approximations still hold.  For simplicity of discussion, we will neglect Helium, since its treatment is completely analogous to the treatment of Hydrogen recombination. We work with units in which the speed of light, $c$, is set to unity.

\subsection{The effective 3-level atom}\label{section:eff3}
Recombination in the early universe is an example of the so called Case B recombination, i.e. all levels with $n\geq 2$ are in radiative equilibrium, but the Lyman $\alpha$ line is optically thick. This means that a $2P\to 1S$ transition generates a high energy photon which immediately causes a $1S\to 2P$ transition in a nearby atom, which results in an overpopulation of the $n=2$ level relative to radiative equilibrium. The recombination to the $1S$ level proceeds in two channels: either through the redshift of Ly$\alpha$ photons out of the absorption line, or through a two-photon decay from the $2S$ level. Thus, in the homogeneous universe one can treat Hydrogen recombination using the effective 3-level atom picture described by \cite{peebles}. Due to its simplicity, it is tempting to use the Peebles equation for recombination in the presence of perturbations. However, the Peebles 3-level atom is self-consistent only because the physical timescales involved in recombination span a huge range of scales. Thus, a number of processes can be regarded as instantaneous allowing one to make several quasi-equilibrium approximations which we will discuss in detail below. In the presence of perturbations, however, one should consider whether the perturbation timescale is long enough to allow for the same approximations used in the homogeneous case. In order to answer this question, we will list the most important approximations together with their relevant timescales, where appropriate, which allow one to calculate the recombination history to an accuracy of about $1\%$ \cite{sasselov}. We will delay the discussion of the approximations allowing for the calculationg of the Ly$\alpha$ escape probability to the next section. As we go along, we will gather all relevant timescales in Figures \ref{fig:thermal} and \ref{fig:recomb}. For reference, the recombination timescale is $t_{rec}\equiv |n_e/\dot n_e|\approx 5\times10^4\,$yr ($=15\,$Mpc in conformal time) during recombination (cf. Fig. \ref{fig:thermal}). 

The first relevant timescale enters in the calculation of the photoionization and recombination cross sections. This calculation relies on the fact that the electrons and baryons obey the Maxwell-Boltzmann distribution with a common kinetic temperature, $T_M$. This is justified beacuse the thermalization timescale is given by the collision timescale, which is always less than about a year for the relevant redshifts ($z>500$), corresponding to a comoving $k\gtrsim 10\,$kpc$^{-1}$ (see Fig. \ref{fig:thermal}) which is way beyond the scales we care about. 
Other processes, such as collision, stimulated emission and stimulated recombination could be in principle important for recombination, but they are not, as shown in  \cite{sasselov}.

\begin{figure}
\begin{center}
%\plotone{f3}
\includegraphics[width=15cm]{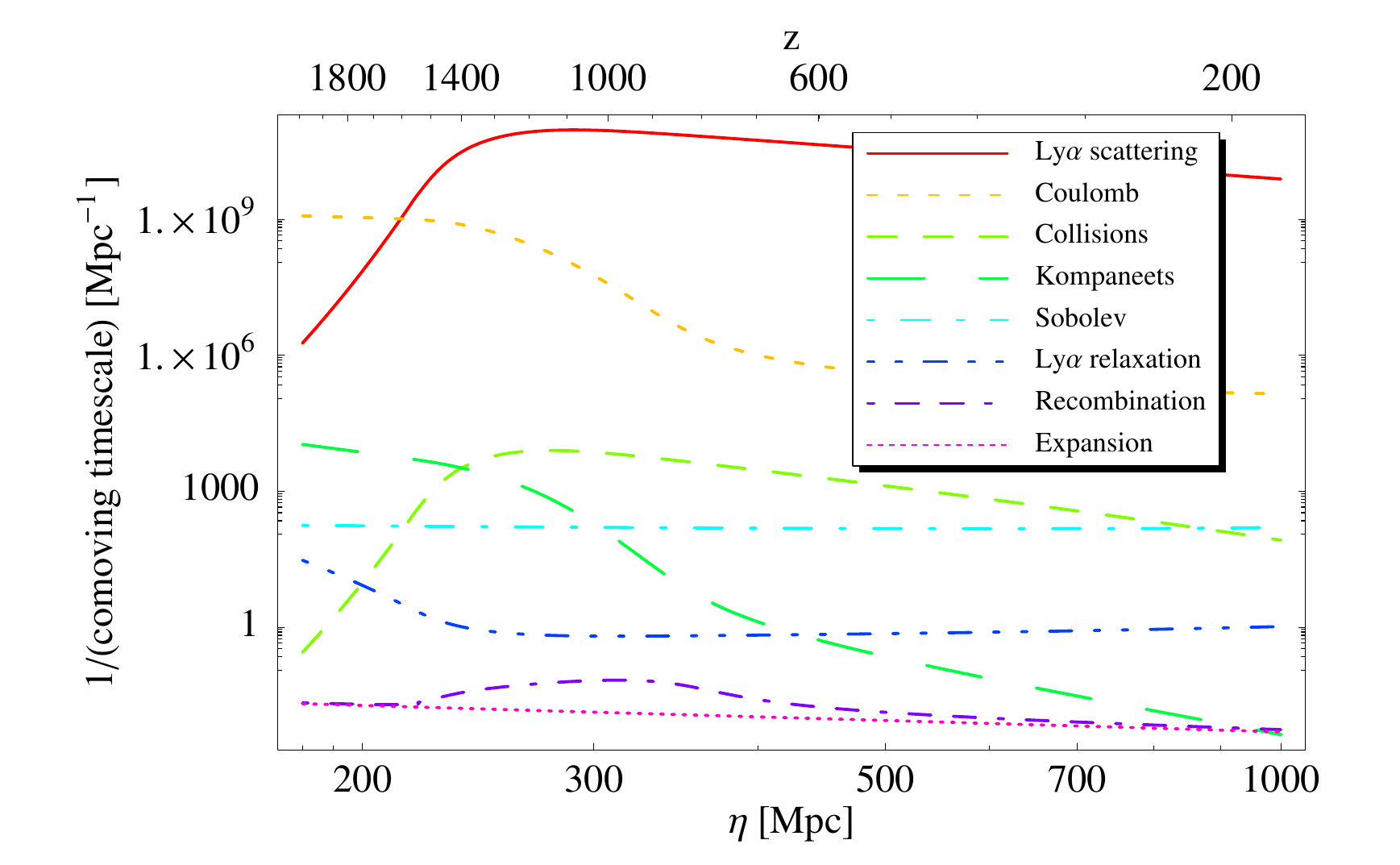}
\end{center}
\caption{We summarize all important timescales related to recombination in this figure and in Fig. \ref{fig:recomb}. Here we concentrate on the timescales that control the 3-level atom approximation. The peak of the CMB visiblity function is at $z=1090$. The expansion timescale is defined as $(3\mathcal{H})^{-1}$; the Ly$\alpha$ relaxation timescale is given by eq.~(29) of \cite{rybdell}; the collisions timescale corresponds to atomic collisions, and is irrelevant at early times; the Kompaneets timescale corresponds to the timescale of Compton heating. Of the timescales given in this figure, the Ly$\alpha$ relaxation imposes an upper limit of $k\approx1\,$Mpc$^{-1}$ on the validity of the perturbed Peebles equation. } \label{fig:thermal}
\end{figure}

%2) Collisions are unimportant \cite{sasselov}. 

%3) Stimulated emission and stimulated recombination are unimportant \cite{sasselov}.

Next, another important assumption is that all levels with $n\geq 2$ are in radiative equilibrium. This comes from the fact that,  apart from the Lyman series, the number of photons generated by atomic transitions can be completely neglected with respect to the number of photons in the Planck spectrum at the corresponding frequencies \cite{sunyaev}. Since the recombination rates involve a collision with an electron, they are orders of magnitude less frequent than radiative transitions, which therefore should keep the populations with $n\geq 2$ in thermal equilibrium with the radiation. Thus, the ratio of the populations of two excited states $i,j\geq 2$ is given by the Boltzmann factor with a temperature given by the radiation temperature $T_R$ entering in the black body distribution:
\begin{eqnarray}\label{ninj}
\frac{n_i}{n_j}=\frac{g_i}{g_j}e^{(E_j-E_i)/k_B T_R}\ ,
\end{eqnarray}
where $g_i$ and $E_i$ are the degeneracy and ionization energy of the $i$-th level, respectively, and $k_B$ is Boltzmann constant. This assumption has been checked in \cite{sasselov}, where it was shown  that the above approximation is valid for levels which are nearby in energy (e.g. $n_{70}/n_{75}$), but for $z\lesssim 1000$ it fails to reproduce the ratio of populations for levels which are well-separated in energy (e.g. $n_{70}/n_2$) due to the photoionization becoming inefficient. Nevertheless, this induces an error only at lower redshifts, $z\lesssim 800$, which amounts to no more than about $10\%$ of the electron fraction $x_e$ since at these redshifts the populations of the highly excited states are very small \cite{sasselov}. This correction can be accounted for in the Peebles 3-level atom by the introduction of a fudge factor to the absorption rate \cite{sasselov} as discussed below.  However, this tells us that the timescale for relaxation among very different states is not much faster than the recombination time-scale at $z\sim 800$.  Fortunately, the photoionization rate grows very rapidly as we increase the redshift (see Fig.~(\ref{fig:recomb})), and therefore, even though we are interested in perturbations whose timescale is faster than recombination, we expect the thermal equilibrium approximation among the states $n\geq 2$ is a good approximation to $\sim10\%$ for momenta $k$ up to order  $1\,$Mpc$^{-1}$.

Given the above approximations, we can now proceed along the lines of \cite{peebles}, and write down the equations for the evolution of the electron number density $n_e$ and the population, $n_2$, of the first excited state:
\begin{eqnarray}
&&\frac{\partial n_e}{\partial t}+3 H n_e=-\Delta R_{c\to 2}\label{dnedt1}\ ,\\
&&\frac{\partial n_2}{\partial t}+3 H n_2=\Delta R_{c\to 2}-\Delta R_{2S\to 1S}-\Delta R_{2P\to 1S}\ ,\label{dn2dt}
\end{eqnarray}
where $H=d\ln a/dt$ is the Hubble parameter and $a$ is the scale factor; $\Delta R_{2S\to 1S}$ is the net rate of $2S\to 1S$ two-photon transitions, $\Delta R_{2P\to 1S}$ is the net rate of $2P\to 1S$ transitions, and $\Delta R_{c\to 2}$ is the net transition rate between the continuum and the first excited state, including transitions which involve intermediate states such as continuum $\to (n>2)\to (n=2)$. Notice that we did not include transitions from $n>2$ to $1S$ because they are exponentially suppressed. Because the upper levels are in radiative equilibrium, one can write $\Delta R_{c\to 2}$ in terms of a total recombination $\alpha_B$ and photoionization $\beta_B$ coefficients as follows \cite{peebles}:
\begin{eqnarray}\label{Rc2}
\Delta R_{c\to 2}=\alpha_B(T_M)n_en_{\mathrm{HII}}-\beta_B(T_R)n_{2S}\ ,
\end{eqnarray}
where $n_{\mathrm{HII}}$ is the  number density of free protons, which after neglecting Helium, reduces to $n_e=n_{\mathrm{HII}}$.
Neglecting stimulated recombination, $\alpha_B$ can only depend on the matter temperature (and negligibly on the number density of electrons \cite{1994MNRAS.268..109H}), since recombination is a collisional process which depends on the kinetic temperature of the electrons $T_M$. The Case B  total recombination coefficient is given in \cite{1994MNRAS.268..109H}, and an adequate fitting formula for $\alpha_B(T_M)$ is given in \cite{1991AA...251..680P}: $\alpha_B(x)=1.14\times 10^{-13} a y^b/(1+c y^d)$ cm$^3$ s$^{-1}$, where $y\equiv x/10^4 K$, $x$ is the temperature in Kelvin degrees, and $a,b,c,d$ are dimensionless constants of order 1 ($a=4.309,\; b=-0.6166,\; c=0.6703,\; d=0.5300$), and $1.14$ is the fudge factor introduced by \cite{sasselov} to account for non-equilibrium populations of the excited states. The total recombination coefficient is given by a sum over the recombination coefficients, $\alpha_i(T_M)$, to each excited state. One can derive a total photoionization coefficient $\beta_B$ from detailed balance considerations (see e.g. \cite{sasselov}). In local thermodynamical equilibrium (LTE), using equilibrium populations denoted with a superscript $eq$ one can write $\beta_in_i^{eq}=n_e^{eq}n_{\mathrm{HII}}^{eq}\alpha_i$, where $\beta_i(T_R)$ is the photoionization coefficient for each excited state. Then, the total photoionization coefficient from the first excited state is given by $\beta_B=\sum_{i>1}\beta_i(T_R) (n_i/n_{2})^{eq}$, which can be combined with the LTE expression and (\ref{ninj}) to give (\ref{betaB}) below. Since both the individual photoionization rates and the population ratios (\ref{ninj}) depend only on $T_R$, the total recombination coefficient can also depend only on $T_R$. For the purposes of comparison with the Peebles 3-level atom, \cite{sasselov} obtains an expression for $\beta_B$ which depends only on $T_M$. During recombination, this is an adequate approximation since Compton scattering keeps $T_M$ equal to $T_R$, even for the faster timescales associated with the perturbations. We explicitly verify this by keeping $T_M$ and $T_R$ distinguished and by solving for both of them. In this case, $\beta_B(T_R)$ is given by:
\begin{eqnarray}\label{betaB}
\beta_B(T_R)=\alpha_B(T_R)e^{-E_2/k_BT_R}\left(\frac{2\pi m_e k_B T_R}{h_P^2}\right)^{3/2}\ ,
\end{eqnarray}
where $h_P$ is Planck's constant and $m_e$ is the electron mass. 

The $2S\to 1S$ net transition rate is given by
\begin{eqnarray}\label{R2s}
\Delta R_{2S\to 1S}=\Lambda_{2S\to 1S} n_{2S}-\Lambda_{1S\to 2S} (T_R) n_{1S} \ ,
\end{eqnarray}
where $\Lambda_{2S\to 1S}=8.22458s^{-1}$ \cite{Goldman}. From detailed balance $n_{1S}^{eq}\Lambda_{1S\to 2S}=n_{2S}^{eq}\Lambda_{2S\to 1S}$, we obtain $\Lambda_{1S\to 2S} (T_R)=\Lambda_{2S\to 1S}e^{-B_{12}/k_B T_R}$, where $B_{12}\equiv  E_1-E_2$.  

We will derive the net rate for the $2P\to 1S$ transition (\ref{Rij}) in the next section. Neglecting stimulated emission, and using the fact that the Lyman $\alpha$ line is extremely optically thick we obtain
\begin{eqnarray}\label{R2p}
\Delta R_{2P\to 1S}=P A_{21}\left[n_{2P}-3n_{1S}e^{-B_{12}/k_BT_R}\right]\ ,
\end{eqnarray}
where $A_{21}$ is the rate of spontaneous emission from $2P\to 1S$, and $P\equiv P_{21}$ is the escape probability for the Ly$\alpha$ photons discussed in the next section. Notice that, if for a moment  we set $P$ to one, the above expression is the standard rate $2P\rightarrow 1S$. As it will become clear in the next section, the multiplication by the probability $P$ takes into account of the fact that just a fraction of these transitions are effective. The rates $R_{2P\to1S}/n_{\mathrm{HII}}$ and $R_{1S\to2P}/n_{\mathrm{HII}}$ per HII are shown in Fig. \ref{fig:recomb}. We can see that Lyman $\alpha$ escape is important only very early on ($z\gtrsim1400$), and that the two-photon decay dominates most of the recombination history.

We can now begin to solve the system of equations (\ref{dnedt1}) and (\ref{dn2dt}).
First, we can neglect the left hand side of (\ref{dn2dt}) since the recombination and Hubble timescales are large compared to the net atomic transitions timescale per excited atom which is about $\Lambda_{2S\to 1S}^{-1}\sim 0.1\,$s for each of the three net rates appearing on the right hand side of (\ref{dn2dt}). This is equivalent to saying that the $n_{2S}$ and $n_{2P}$ reservoirs are in quasi-equilibrium \cite{mukh}. In a perturbed universe, this approximation is again justified, since it corresponds to neglecting the perturbation timescale with respect to the net atomic transitions timescale. Assuming equilibrium populations of $n_{2P}$ and $n_{2S}$ due to fast radiative transitions gives $n_2=4n_{2P}/3=4n_{2S}$. Thus, combining (\ref{dn2dt}),(\ref{Rc2}),(\ref{betaB}),(\ref{R2s}) and (\ref{R2p}), we obtain an algebraic equation which can be solved for $n_2$, obtaining:
\bee\label{eq:nepeebles}
n_2=4 \frac{\alpha_B n_e^2+\left(\Lambda_{2S\rightarrow 1S }+3 P A_{21}\right) e^{-\frac{B_{12}}{k_B T_R}}(n_b-n_e)}{3PA_{21}+\Lambda_{2S\rightarrow 1S }+\beta_B}\ ,
\eee
where $n_b$ is the number density of protons  (both in ionized and atomic Hydrogen).
Plugging this solution for $n_2$ into (\ref{dnedt1}), we obtain the standard Peebles 3-level result, with the distinction between $T_R$ and $T_M$ made clear:
\begin{eqnarray}\label{N0first}
&&\frac{\partial n_e}{\partial t}+3Hn_e= Q\left[n_e,n_b,T_M,T_R,H\right]\ , 
\end{eqnarray}
where 
\bee
Q =-[\alpha_B(T_M)n_e^2-\beta_B(T_R)(n_b-n_e)e^{-B_{12}/k_BT_R}]C_H\ , \label{Qhom}
\eee
and
\bee
C_H= 1- \frac{\beta_B(T_R)}{3PA_{21}+\Lambda_{2S \to 1S}+\beta_B(T_R)}\ .
\eee
As we will show in the next section,
\be\label{eq:PA}
 3PA_{21}\approx \frac{8\pi H\nu_{21}^3}{n_b-n_e}\ , 
 \ee
where $\nu_{21}$ is the frequency of the Ly$\alpha$ photons.  Here and in eq.~(\ref{eq:nepeebles}) we have approximated $n_{1S}\approx n_b-n_e$, since even in the presence of the Ly$\alpha$ bottleneck, the ground state population dominates the rest of the excited states. Note that in the above derivation we could not neglect the left hand side of (\ref{dnedt1}) as we did in (\ref{dn2dt}), since the net atomic transitions timescale per electron is $\Lambda_{2S\to 1S}^{-1}\times n_e/n_2\sim t_{rec}$ (see Fig. \ref{fig:recomb}). Thus, in the presence of perturbations, the left hand side of (\ref{dnedt1}) will receive non-negligible corrections, as we will show in Section \ref{section:firstorder}.

\begin{figure}
\begin{center}
%\plotone{f3}
\includegraphics[width=15cm]{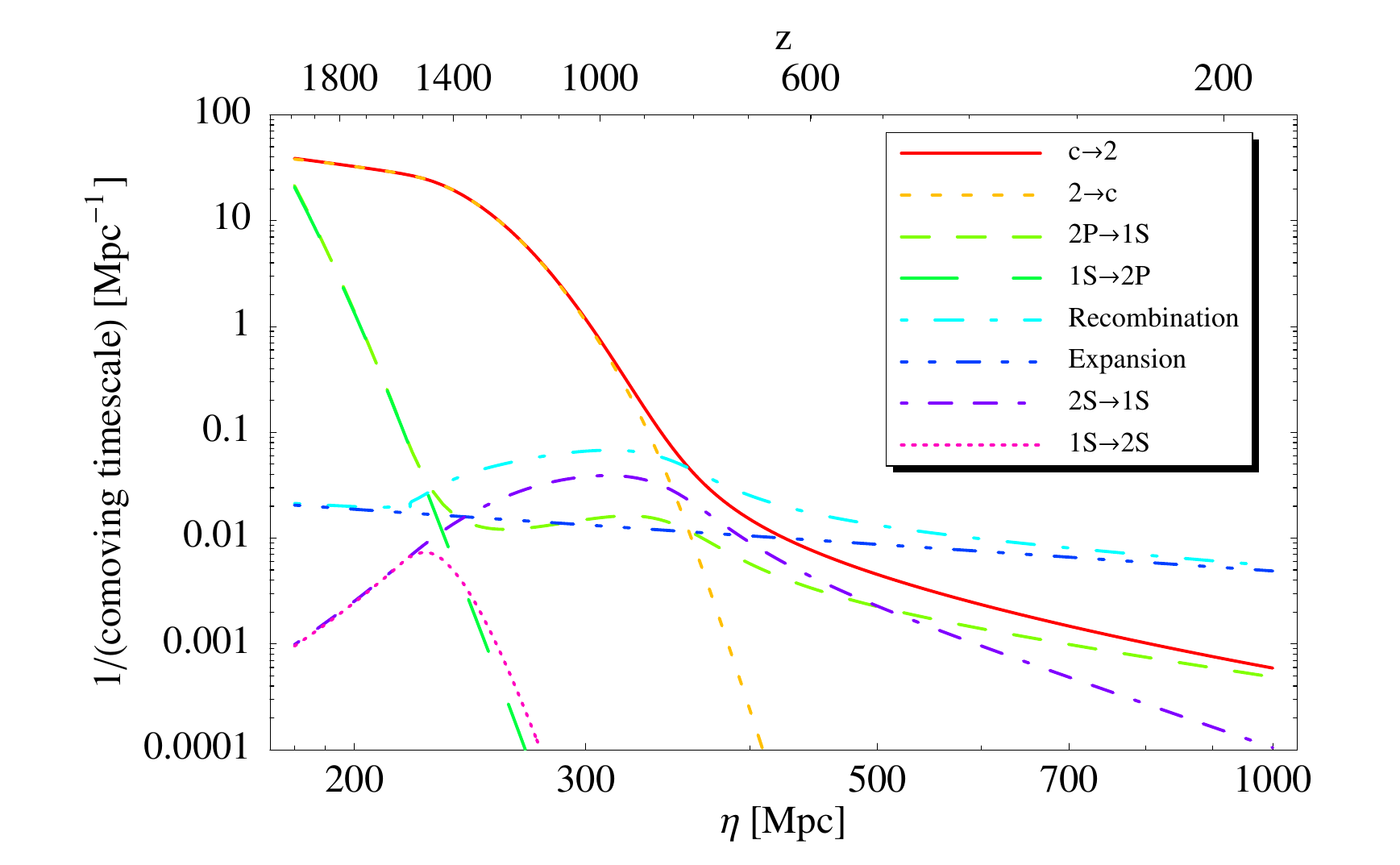}
\end{center}
\caption{The comoving rates per HII ion governing Hydrogen recombination. The recombination rate is defined as   $|\dot n_e/n_e|$. Here we concentrate on the rates that affect directly the recombination rate.} \label{fig:recomb}
\end{figure}

%The above result, (\ref{N0first}), is sufficient to proceed with perturbing the electron density fraction without rigorously deriving the perturbations to the escape probability. However, a rigorous derivation of the perturbations to the escape probability would require a discussion of a series of approximations, including the Sobolev approximation, which are treated in the rest of Section \ref{section:review}. 

\subsection{Treatment of Ly$\alpha$ photons}\label{section:lyhom}
In the expanding universe photons are ``advected'' in momentum space by the cosmological redshift. The photons blueward of the Ly$\alpha$ line are redshifted into the line, where they diffuse until they reach the red wing of the line from which they can ``escape''.  In this section we derive the Boltzmann equation for the evolution of the photon distribution function. Solving that equation will allow us to find the escape probability of Ly$\alpha$ photons and derive eq.~(\ref{R2p}). 

In general, the physical photon number density in each frequency bin of width $\delta\nu$ is given by $8\pi\nu^2F_\gamma(\nu(t),t)\delta\nu$, where $8\pi\nu^2$ gives the number of states per unit volume per unit frequency range, and $F_\gamma(\nu,t)$ is the photon phase-space distribution (one can think of it as the occupation number of each state). Note that we have included the photon degeneracy in $8\pi\nu^2$. Following a given frequency bin as it is being redshifted, the number of states in a comoving volume $V$ is given by $a^3(t) 8\pi V \nu(t)^2\delta \nu(t)$, which is constant in time. Therefore, the rate at which the number density of photons increases due to photons which are redshifted into a bin of a given fixed frequency $\nu$ between a time $t-\delta t$ and $t$ is simply given by $$R_{in}=8\pi \nu^2\delta \nu F_\gamma[\nu (1+H\delta t),t-\delta t]/\delta t\ ,$$ where we used $\delta\nu/\delta t=-H\nu$. The rate at which photons are redshifted out of the bin is instead given by $$R_{out}=8\pi \nu^2\delta \nu F_\gamma[\nu,t]/\delta t\ .$$ The net flux divergence equals therefore
\begin{eqnarray}\label{netfluxdiv}
R_{in}-R_{out}&=&-8\pi \nu^2 \left(\frac{\partial F_\gamma}{\partial t}-H\nu\frac{\partial F_\gamma}{\partial \nu}\right)\delta \nu=-8\pi\nu^2 \frac{dF_\gamma}{dt}\delta \nu \ .
\end{eqnarray}
The above expression should equal the net rate at which photons are being absorbed in a certain line representing a transition from the state $j$ to the state $i$: 
\be\label{coll_abs}
\hbox{net absorption rate}=-[A_{ji}n_j-(B_{ij}n_i-B_{ji}n_j)2\nu^3F_\gamma(\nu)]\phi(\nu)\delta \nu\ ,
\ee
where $A_{ji},B_{ji}$ and  $B_{ij}$ are the standard Einstein coefficients, and $\phi(\nu)$ is the line profile \footnote{In the photon collision term in the above eq.~(\ref{coll_abs}) we assumed that complete redistribution holds by setting the emission and absorption profiles equal. In reality this is far from true in general processes. To see this, consider a 2-level atom with levels 1 and 2, and for simplicity let us concentrate on the Ly$\alpha$ transition (i.e. level 1 is sharp) which is the most important during recombination. When an atom is excited to level 2, it spontaneously emits a photon within $\sim 10^{-9}\,$s. During this whole process we can assume that the atom moves with constant velocity, since collisions and atom recoil can be neglected. In its rest frame, the atom emits a photon with the standard Lorentzian profile. This is because we can assume that the absorbed and emitted photons are not correlated in the atom rest frame since the absorption and emission events are well separated in time. However, the atoms follow a Maxwell-Boltzmann distribution, which implies that in the laboratory rest frame, the absorbed and emitted photons are correlated, with the correlation given by the so-called redistribution function $R(\nu',\hat{n}';\nu,\hat{n})$ which depends on the frequencies of the incoming and outgoing photons and on their directions. Neglecting angular redistribution \cite{humryb}, this results in an emission profile given by $\phi_{21}(\nu)=\int d\nu' I(\nu')R_{\mathrm{II}}(\nu',\nu)/\int d\nu' I(\nu')\phi_{12}(\nu')$, where $R_{\mathrm{II}}(\nu',\nu)$ is the partial frequency redistribution function subject to the above set of assumptions \cite{hummer}. To understand how one obtains the redistribution function for an $n$-level atom, one should start from the formulation of the radiative transfer problem in terms of quantum mechanical generalized redistribution functions. For a systematic discussion of how this is done, see \cite{HOSI} and \cite{HOSII}. Using the correct redistribution function $R_{\mathrm{II}}$ for the Ly$\alpha$ transition, the radiative transfer equation can be parametrized in the form given in \cite{humryb} for example. It was shown in \cite{humryb} that the line profile averaged intensity, which is what enters the net transition rate, is weakly dependent on the choice of the redistribution function for $\gamma$ (defined below in eq.~(\ref{gammaS})) in the range $10^{-4}$ to $1$. Outside this range, the result using complete redistribution (corresponding to $R_{\mathrm{CR}}=\phi(\nu)\phi(\nu')$) is practically identical to the one using the $R_{\mathrm{II}}$ redistribution function. For the Ly$\alpha$ photons $\gamma$ is of order $\sim 10^{-9}/(1-x_e)$ during recombination. Thus, once we have non-negligible recombination, we can work with the complete redistribution approximation. }. In the above equation we used that the intensity of the radiation field is given by $I=2\nu^3F_\gamma$, and that collisions are unimportant during recombination as shown by \cite{sasselov}. Equating the net absorption rate to the flux divergence we obtain the following differential equation
\begin{eqnarray}\label{photonBoltz}
\nu\frac{d F_\gamma}{d t}=\frac{1}{8\pi\nu}[A_{ji}n_j-2\nu^3(B_{ij}n_i-B_{ji}n_j)F_\gamma]\phi(\nu) \ .
\end{eqnarray}
This is exactly the Boltzmann equation for the photon distribution. In App.~\ref{section:A} we review the covariant formulation of the Boltzmann equation necessary for perturbing it. 

Before actually solving this equation, we can already  estimate the rate $\Delta R_{2P\to 1S}$, which must equal $R_{out}-R_{in}$. We can approximate eq.~(\ref{netfluxdiv}) by $\Delta R_{2P\to 1S}\approx 8\pi H\nu^3[F_\gamma(\nu_R)-F_\gamma(\nu_B)]$ with $\nu_R$ and $\nu_B$ two frequencies just to the red and blue of the line. Just blueward of the line, the spectrum converges to the black body spectrum $B_\nu$, i.e. $2\nu_B^3F_\gamma(\nu_B)=B_\nu$. For optically thick lines, the photon distribution function saturates to $S/2\nu^3$ at $\nu_R$, with $S$ being the standard radiative transfer source function (see (\ref{SourceF}) below). Thus, we can write $\Delta R_{2P\to 1S}\approx 4\pi H(S-B_\nu)$, which as we will show below (cf. equations (\ref{PjiFINAL}) and (\ref{Rij})) is an excellent approximation for optically thick lines. By using the standard relation $A_{ji}=g_\gamma \nu^3 B_{ji}$, it is easy to see that this result is equivalent to eq.~(\ref{R2p}) that we used in Section \ref{section:eff3}.

Let us now proceed and solve the Boltzmann equation (\ref{photonBoltz}) to obtain the rate $\Delta R_{c\rightarrow 2}$. Following \cite{rybicki} we can reparametrize (\ref{photonBoltz}) using the proper frequency $\nu(t)$ of an individual photon as a time parameter, since $\nu(t)\propto a(t)^{-1}$ is monotonous:
\begin{eqnarray}\label{Fdiff}
 -\gamma\frac{dF_\gamma}{d\nu}& =&\left[\frac{S}{2\nu^3}-F_\gamma\right]\phi(\nu) \ ,
\end{eqnarray}
where
\begin{eqnarray}\label{gammaS}
 \gamma \equiv  \frac{4\pi H}{B_{ij}n_i-B_{ji}n_j}\,  ,
\end{eqnarray}
is usually referred to as the Sobolev parameter, and
\begin{eqnarray}\label{SourceF}
S\equiv\frac{A_{ji}n_j}{B_{ij}n_i-B_{ji}n_j} \ ,
\end{eqnarray}
is usually called the source function.
The solution to (\ref{Fdiff}) is given by
\begin{eqnarray}\label{inttrans}
&&F_\gamma(\nu)=F_\gamma(\nu_B)e^{-\tau(\nu)}+e^{-\tau(\nu)}\int_{\nu_B}^{\nu}e^{\tau(\nu')}\frac{S(\nu')}{2\nu'^3}\frac{d\tau(\nu')}{d\nu'}d\nu'\ , 
\end{eqnarray}
with
\be\label{eq:taunu}
\tau(\nu)\equiv  \int_{\nu}^{\nu_B}\frac{\phi(\nu')}{\gamma(\nu')} d\nu' \ .
\end{eqnarray}
The optical depth $\tau(\nu)$ is calculated along incoming rays. We will see later that the final result is independent of whether one considers incoming or outgoing rays. The choice of $\nu_B$ is arbitrary, but for the approximations below to work, we choose it to be just blueward of the line. The frequencies just  blueward of the line are not affected by the line, and  therefore, we can write $F_\gamma(\nu_B)=F_{bb}(\nu_B)\approx F_{bb}(\nu_{ij})$, where $F_{bb}$ is the Bose-Einstein (black body spectrum divided by $2\nu^3$) distribution. 

The net transition rate from $j\to i$ is given by
\begin{eqnarray}\label{Rtemp2}
\Delta R_{ji}=\int\limits_0^\infty  d\nu\phi(\nu)[A_{ji}n_j-2\nu^3(B_{ij}n_i-B_{ji}n_j)F_\gamma]\ .
\end{eqnarray}
The integrand is multiplied by the line profile which is centered around $\nu_{ij}$ and has a width $\Delta \nu$ given by the thermal broadening of the line. Since $\nu$ plays the role of a time coordinate, in configuration space this implies that the contribution of the number densities entering the integrand will be only from a spacetime patch of scale $L_S=v_{\mathrm{thermal}}/H$, which we call the Sobolev scale~\footnote{For a single photon, the Sobolev patch extends a scale $L_S$ in the time direction, and a much smaller scale in the space direction, since the photon experiences a random walk. The relevant scale is given by $L_S$, the largest of the two.}. Here $v_{\mathrm{thermal}}$ is the average thermal speed of the atoms. We can take the Sobolev scale as approximately constant during recombination and equal to $L_S=6.3$ kpc ($\approx20\,$yr in physical time). This is very small compared to the recombination timescale (which is also the timescale of variation of the source function), about $15\,$Mpc. 

This allows us to make the so called large velocity gradient or Sobolev approximation \cite{sobolev} where we neglect the spatial and temporal variations of the number densities over the Sobolev scale.
This is not an innocuous approximation: this amounts to neglecting the relaxation time over which  $F_\gamma$ becomes quasi-static in the line. The validity of this approximation has been checked by \cite{rybdell}, where they show that the relaxation timescale of $F_\gamma$, defined in \cite{rybdell} as the timescale over which the intensity in the line reaches within $10\%$ of the quasi-static value, is $t_{\rm rell}\sim10^{3.5}\,$yr in physical time ($t_{\mathrm{rell}}^{-1}\approx1\,$ comoving Mpc$^{-1}$) at the peak of the visiblity function (see Fig. \ref{fig:thermal}). This effect, which becomes relevant for modes of order $1\,$Mpc$^{-1}$, is not very important because the transition 2P $\rightarrow$ 1S is also not very important~\footnote{The effect of the Ly$\alpha$ escape probability is about 10\% for $k\simeq0.1$ Mpc$^{-1}$ (equivalent to a CMB multipole $l\sim 1500$) and about 25\% for $k\simeq0.2$ Mpc$^{-1}$ ($l\sim 3000$).}.  

Neglecting variations of the source function and the Sobolev parameter we can rewrite (\ref{Rtemp2}) as
\begin{eqnarray}\label{Rtemp1}
\Delta R_{ji}=A_{ji}n_j-(B_{ij}n_i-B_{ji}n_j)2\nu_{ij}^3 \int\limits_0^\infty F_\gamma \phi(\nu) d\nu\ .
\end{eqnarray}
In the same approximation we can take $\gamma$ outside of the integral in (\ref{eq:taunu}), and obtain $\tau(\nu)=\gamma^{-1}(\nu_{ij})\int_\nu^{\nu_B} d\nu'\phi(\nu')$. At this point we are allowed to extend the integral range to $\nu_B\to\infty$, since the contribution of the line profile at these frequencies is negligible. Thus, we can finally write
$$\tau(\nu)=\gamma^{-1}(\nu_{ij})\int^{\infty}_\nu d\nu'\phi(\nu')\ ,$$ in the Sobolev approximation. The integral in (\ref{Rtemp1}) can be evaluated using (\ref{inttrans}) by a change of variables using $d\tau=-\gamma^{-1}\phi(\nu)d\nu$. In the end we obtain 
\begin{eqnarray}\label{eq:intensity}
2\nu_{ij}^3\int\limits_0^\infty F_\gamma \phi(\nu) d\nu=P_{ji}2\nu_{ij}^3F_{bb}(\nu_{ij})+(1-P_{ji})S(\nu_{ij})\ , 
\end{eqnarray}
where
\begin{eqnarray}\label{RijTemp111}
P_{ji}\equiv \int^\infty_0d\nu\phi(\nu)\exp[-\tau(\nu)]= \gamma\left(1-e^{-1/\gamma}\right)\ .\label{PjiFINAL}
\end{eqnarray}
The prefactor $P_{ji}$ is called the escape probability, as it is the probability for a line photon to escape to infinity without scattering, once it is spontaneously emitted. To see why this is so, let us consider the different components of the line-averaged intensity given by the above equation (\ref{eq:intensity}). The first term, $2\nu_{ij}^3P_{ij}F_{bb}(\nu_{ij})$, is called the direct component, since this is the intensity coming from the blue part of the spectrum which equals $2\nu_{ij}^3F_{bb}(\nu_{ij})$ times the probability for it to ``escape'' to the center of the Sobolev patch. The second term is called the diffuse component of the radiation field, $(1-P_{ji})S(\nu_{ij})$, since it is that part of the intensity which is locally produced, and which does not escape the Sobolev patch. Using the definition of the source function, we can write
\begin{eqnarray}\label{dich}
n_jA_{ji}+[n_jB_{ji}-n_i B_{ij}](1-P_{ji})S=n_jA_{ji}P_{ji}\ ,
\end{eqnarray}
%In terms of the separation of the radiation field into components, called the dichotomous model \cite{rybicki}, we can intepret the above equation as follows: 
which means that the net rate of producing local (diffuse) photons equals the rate of spontaneously emitting escaping photons. Using (\ref{SourceF}), (\ref{Rtemp1}) and (\ref{RijTemp111}) we can finally write
\begin{eqnarray}\label{Rij}
\Delta R_{ji}=P_{ji}[A_{ji}n_j-(B_{ij}n_i-B_{ji}n_j)2\nu_{ij}^3 F_{bb}(\nu_{ij})]\ .
\end{eqnarray}
Neglecting spontaneous emission, and specializing to the Ly$\alpha$ transition, we recover (\ref{R2p}) as wished.
We notice that in the Sobolev approximation, after averaging over the line profile (i.e. integrating by $\int d\nu \phi(\nu)$), eq.~(\ref{inttrans}) becomes:
\be\label{eq:simpleF}
\bar F_\gamma=P F_{bb}+(1-P)\frac{S(\nu_{ij})}{2\nu_{ij}^3}\ ,
\ee
where the bar stays for frequency averaging. We will find a very analogous expression in App.~\ref{section:B} when we will generalize to the perturbed universe.

As anticipated, we also notice that the Sobolev approximation ensures that the forward and backward escape probabilities are equal, since 
\bee
P_{ji}= \int^\infty_0d\nu\phi(\nu)\exp\left[-\gamma^{-1}(\nu_{ij})\int^{\infty}_\nu d\nu'\phi(\nu')\right]=\int^\infty_0d\nu\phi(\nu)\exp\left[-\gamma^{-1}(\nu_{ij})\int^{\nu}_{0} d\nu'\phi(\nu')\right]\ . \nonumber
\eee

In the discussion above we neglected the secondary distortions to the radiation spectrum \cite{sasselov}. These distortions come from the redshifting of line photons into a neighbouring line. This effect is trying to invalidate the approximation  $F_\gamma(\nu_B)\simeq F_{bb}(\nu_B)$ that we used in the derivation of the escape probability. This approximation holds everywhere outside the Lyman series, where the relative intensity distortion is $\Delta I/(2\nu^3F_{bb})\lesssim 10^{-6}$ \cite{sunyaev}. The secondary distortions due to the higher-energy Lyman lines on the lower-energy Lyman lines has been assessed in \cite{astro-ph/0702531v2}. The biggest effect comes from the secondary distortion caused by the Ly$\beta$ line on the Ly$\alpha$ line. However, the number of extra Ly$\alpha$ photons per H atom due to this effect is $0.2\%$ \cite{astro-ph/0702531v2}, therefore the effect of perturbations on the number of redshifted Ly$\beta$ photons will have negligible impact on $\delta_e$.
%  The maximum ratio $\Delta I/(2\nu^3F_{bb})= 64\%$ due to Ly$\beta$ is achieved at $z=1450$, and those extra photons reach Ly$\alpha$ at $z=1220$ \cite{astro-ph/0702531v2}.
%However, for $\gamma\to 0$, the net transition rate is given by $\Delta R_{ij}=4\pi H(S-2\nu_{ij}^3F_{\gamma}(\nu_B))$. Thus, the effect of the term from the blue part of the spectrum can be neglected for Ly$\alpha$ since $S\gg 2\nu_{21}^3F_\gamma(\nu_B)$ for the Ly$\alpha$ line \cite{astro-ph/0702531v2}.

This concludes our discussion of Hydrogen recombination in the unperturbed universe and of the escape probability. More details on the Sobolev escape probability can be found  \cite{rybicki}, while  a recent review on of recombination is given in \cite{arXiv:0711.1357v1} and the references therein. 

Throughout this discussion so far we have neglected Helium recombination. We can easily reintroduce it: in this case, the Peebles equation for $n_e$ becomes an equation for $n_{\mathrm{HII}}$, and one ends up with two extra equations for the abundance of HeII and HeIII, given in a convenient form for example in \cite{Novosyadlyj}~\footnote{Note that we do not need to revert to the more detailed equations given in \cite{arXiv:0711.1357v1}, because Helium recombination has little effect on the CMB.}.

\section{Perturbing recombination 
%first order equations for $\delta_e$ and $\delta T_M$
}\label{section:firstorder}

We are now ready to derive the equations for $\delta_e$ and $\delta T_M$. Throughout the rest of this paper, we will work in a spatially flat FRW background. We work with a metric with positive signature. Greek letters denote spacetime indices; latin letters denote spatial indices. In this section, we will work to first order in the metric perturbations in the Newtonian gauge in which the line element is given by
\begin{eqnarray}
ds^2=a^2(\eta)[-(1+2\Psi)d\eta^2+(1-2\Phi)dx^2]\ ,
\end{eqnarray}
where $\eta$ is conformal time, $\Psi$ and $\Phi$ are the Newtonian gauge potentials, and $a(\eta)$ is the scale factor. A dot will denote a partial derivative with respect to $\eta$.

\subsection{Electron density perturbations\label{sec:first_order_eq}}
In the previous section we found that all the approximations entering in the Peebles 3-level atom equation (\ref{N0first}) are still valid in the presence of perturbations with $k<1\,$Mpc$^{-1}$. This limit comes both from the timescale over which the excited states reach thermal equilibrium, and from the timescale over which the photon distribution reaches quasi-equilibrium in the Ly$\alpha$ line. Limiting ourselves to $k$'s lower than this, we are therefore going to study the Peebles 3-level atom in a perturbed universe. In order to do this, we have to derive the escape probability in the presence of metric perturbations, and we also have to write down what the Boltzmann equation for the electrons is in curved spacetime. The last step is really necessary because, contrary to the treatment of first order perturbations, in the case of recombination the collision term does not vanish in the homogeneous universe. This means that we have to consistently treat the metric perturbations both in the Liouville (or free-streaming) term, and in the collision term.

We start from the covariant form of the Boltzmann equation. The covariant formulation of the Boltzmann equation in a curved spacetime is reviewed in App.~\ref{section:A}, to which we will refer on several occasions in order to use some results from covariant kinetic theory. From (\ref{boltz1}) in App.~\ref{section:A} the Boltzmann equation can be written as
\begin{eqnarray}\label{Boltzmann}
P^0\frac{d f}{d\eta}=P^\mu\frac{\d\, f}{\d x^\mu}+\frac{d P^i}{d\lambda}\frac{\d\, f}{\d P^i}=C[f]\ ,
\end{eqnarray}
where $f(x^i,P_j;\eta)$ is the one-particle distribution function; $P^\mu= dx^\mu/d\lambda$
 where $\lambda$ is the affine parameter chosen so that $P^\mu$ is the momentum of the particle ($d \lambda$ coincides with the time interval of a local inertial observer divided by the energy observed in the same frame: $d\lambda=dt_{\rm inertial}/E_{\rm inertial}$), $C[f]$ is the collision term given in (\ref{C}) in App.~\ref{section:A} . The one-particle distribution function $f(x^\mu,P^\nu,\eta)$ is a scalar function of the phase-space variables, which gives the distribution of particle momenta at every spacetime point. A very simple way to understand the structure of the Boltzmann equation is to notice that the left hand side  of eq.~(\ref{Boltzmann}) is just the total derivative of the distribution function with respect to the affine parameter $\lambda$. Since these are scalar quantities, the collison term $C[f]$ on the right hand side is the same as the one measured in a local inertial. The factor of $P^0$ is due to the gravitational redshift in converting the coordinate time to the time of the local inertial frame where $C[f]$ gets a very simple form (see~App. \ref{section:A}). In fact, it is very useful to  express the collision term at a certain given coordinate $x^\mu$ using the momenta $p^i$ of the particles measured in the Local Inertial Frame Instantaneously at Rest with respect to the Comoving Observer (LIFIRCO) at fixed $x^i$. In this case the cross sections have the same expressions as in Minkowski space, and the collision term contains no metric fluctuations. All the metric fluctuations are confined to the Liouville term.  This leads  us to often use mixed coordinates $(x^i,p^j)$, which we call ``nice'' coordinates, where the spacetime coordinates are generic, for example they can be the ones associated to the Newtonian gauge, while momenta are the ones corresponding to the LIFIRCO. The distribution function expressed with these nice coordinates we denote with $F(x^i,p^j;\eta)\equiv f(x^i,P_k(x^i,p^j;\eta );\eta)$. 

From the Boltzmann equation (\ref{Boltzmann}), one can derive the conservation of 4-momentum for a given particle species given in (\ref{coll}) in App.~\ref{section:A}: 
\be\label{eq:divergency}
(n\, U^\mu)_{;\mu}=\int \pi\; C[f]\ ,
\ee 
where $n$ is the number density of particles measured in a local inertial frame momentarily at rest with respect to the fluid.
%(to first order, $n$ equals the number density measured by comoving observers)
 $\pi$ is the invariant measure in momentum space on the mass-shell, which reduces to $g_{\mathrm{deg}}d^3p/E$ in a local inertial frame (see eq.~(\ref{pi}) in App.~\ref{section:A}), where $E\equiv p^0$ is the particle energy as measured in that frame, and $g_{\mathrm{deg}}$ is the number of internal degrees of freedom for each particle of the given species. We have defined the 4-velocity $U^\mu$ in such a way that the particle current ${\cal{N}}^\mu$ can be expressed in the following form $${\cal{N}}^\mu=n\, U^\mu\ .$$We can write equation~(\ref{eq:divergency}) in nice coordinates:
\begin{eqnarray}\label{nmu}
(n\, U^\mu)_{;\mu}=g_{\mathrm{deg}}\int\frac{d^3 p}{E}C[F]\ .
\end{eqnarray}
The Newtonian gauge 4-momentum $P^\mu$ is related to the 4-momentum $p^\mu=(E,p^i)$, measured in the LIFIRCO, via the tetrad (\ref{tetrad_P}) discussed in App.~\ref{section:A}. We call the 3-velocity of the fluid measured in the LIFIRCO as $v^i=p^i/E$. At first order, the relation between $U^\mu$ and $v^i$ can be found in e.g. \cite{maed}, or by using (\ref{4vel}), which we will derive later. To first order this is given by:
\begin{eqnarray}\label{U}
U^0=a^{-1}(1-\Psi)\ , \quad U^i=a^{-1}v^i\ .
\end{eqnarray}

For a nuclei species $n$, number conservation holds (i.e. the integrated collision term in (\ref{nmu}) vanishes in this case), since nucleosynthesis is long over by the time of recombination. Therefore, from (\ref{nmu}) and (\ref{U}) we have
\begin{eqnarray}
&&\dot n_n^{(0)}+3n_n^{(0)}\mathcal{H}=0 \label{num_cons_0}\ ,\\
&&\theta_n+\dot\delta_n-3\dot\Phi=0 \label{num_cons_1}\ ,
\end{eqnarray}
where we have expanded $n_n= n_n^{(0)}(1+\delta_n)$, and we have defined $\theta_n\equiv v^i_{n,i}$ (this coincides at first order to the  $\theta$ variable in \cite{maed}). The above equations are the usual continuity equations written in the Newtonian gauge (see e.g. \cite{maed}).

For the electrons the collision term does not vanish, and we obtain, from (\ref{nmu}):
\begin{eqnarray}
&&\dot n_e^{(0)}+3\mathcal{H}n_e^{(0)}=aQ^{(0)} \label{N_e_0}\ ,\\ \label{eq:delta_e_first}
&&n_e^{(0)}(\dot \delta_e-\dot\delta_b)=a(\Psi Q^{(0)}+\delta Q)-\delta_e a Q^{(0)} \label{N_e_1}\ ,
\end{eqnarray}
where $\mathcal{H}\equiv\dot a/a=a\, H$.  In deriving (\ref{N_e_1}) we used the baryon number conservation equation (\ref{num_cons_1}) to express $\dot\Phi$ in terms of the perturbation to the baryon number density $\delta_b$. We have split the integrated recombination collision term $Q$  into a zeroth and a first order part as 
\bee\label{eq:new_collision}
Q\left[n_e^{(0)}(1+\delta_e),n_b^{(0)}(1+\delta_b),T_M^{(0)}+\delta T_M,T_R^{(0)}+\delta T_R,H^{(0)}(1+\delta_H)\right]= Q^{(0)}+\delta Q\ .
\eee
Notice that, as we have explained before and we explain more in detail in App. \ref{section:A}, the collision term, being a local and scalar quantity, once expressed in terms of nice momenta $p^i$'s does not contain metric perturbations. This means that it is the same function as in the homogeneous case (eq.~(\ref{N0first})), expressed in terms of the perturbed quantities $n_e^{(0)}(1+\delta_e),\,n_b^{(0)}(1+\delta_b),\,T_M^{(0)}+\delta T_M$, and $T_R^{(0)}+\delta T_R$. The only exception to this reasoning is the escape probability, which, as we saw in the former section, is rather a non local quantity. However, for perturbations with wavelength much longer than the Sobolev length, we can imagine to recover the expression for the perturbed escape probability from the result in the homogeneous case. In fact, by inspection of eq.~(\ref{eq:PA}}), we see that what controls the escape probability is the Hubble rate $H$,  which in a homogeneous universe is exactly equal to the local baryon velocity divergence, $U^\mu_{b\ ;\mu}/3=H$. This is not a coincidence, as in fact what controls the probability for a photon to escape is the redshift of the neighboring atoms, which comes from their relative velocity. It is at this point  easy to guess what is the expression for the escape probability (and therefore for the integrated collision term $Q$ above), in the case of a perturbed universe:  in the expression for $Q$, we need just to replace $H$ with $U^\mu_{b\ ;\mu}/3$, or equivalently, as already expressed in eq.~(\ref{eq:new_collision}), $H$ with  $H(1+\delta_H)$, where 
\begin{eqnarray}\label{delta_K_P}
\delta_H=\frac{U^\mu_{b\ ;\mu}}{3H}-1=-\Psi-\frac{\dot\delta_b}{3\mathcal{H}}\ .
%-\Psi-\frac{\dot\delta_b}{3\mathcal{H}}+\frac{1}{3\mathcal{H}}\xi\left(c_s^2k^2\delta_b+\frac{4}{3}\frac{\rho_\gamma}{\rho_b}an_e\sigma_T(\theta_\gamma-\theta_b)\right)\approx
\end{eqnarray}

%The expression for the integrated collision term can be read off by comparing (\ref{N_e_0}) and (\ref{N0first}) from which we see that the above $Q$ is identical to the one in (\ref{Qhom}). We have incorporated the perturbation to the escape probability due to the local velocity divergence in $\delta_H$. In deriving (\ref{N_e_1}) we used the baryon number conservation equation (\ref{num_cons_1}) to express $\dot\Phi$ in terms of the perturbation to the baryon number density $\delta_b$. 

%There are two unknown quantities in the equation for $\delta_e$, namely $\delta_H$ and $\delta T_M$. One can find $\delta_H$ by inspecting the integrated recombination collision term given in (\ref{Qhom}). What enters there is the escape probability, which depends on $H$, which in a homogeneous universe is exactly the local baryon velocity divergence, $U^\mu_{b\ ;\mu}/3=H$. Thus, we automatically obtained a covariant expression for this term, which gives
%\begin{eqnarray}\label{delta_K_P}
%\delta_H=\frac{U^\mu_{b\ ;\mu}}{3H}-1=-\Psi-\frac{\dot\delta_b}{3\mathcal{H}}
%-\Psi-\frac{\dot\delta_b}{3\mathcal{H}}+\frac{1}{3\mathcal{H}}\xi\left(c_s^2k^2\delta_b+\frac{4}{3}\frac{\rho_\gamma}{\rho_b}an_e\sigma_T(\theta_\gamma-\theta_b)\right)\approx
%\end{eqnarray}
By using these arguments, the above expression for the perturbed escape probability had been derived in App. D in \cite{astro-ph/0702600v2}. A rigorous derivation of the perturbations to the escape probability is given in our  App. \ref{section:B} by angle averaging the photon distribution function using the Boltzmann equation. We show that there are extra terms that appear in the expression for $\delta_H$ due to the change of the baryon fluid velocity due to the acceleration by pressure forces and Thomson drag over a time interval $\sim L_S$, but we argue that those contributions are negligible. This gives a rigorous, though more convoluted, derivation of  eq.~(\ref{delta_K_P}).

So far we have neglected Helium recombination, which can be introduced by complete analogy. The perturbations to the HeIII density are given by perturbing the Saha equation, since HeIII recombination procedes very nearly perfect local thermal equilibrium. The equations for the HII and HeII fractional perturbations are given by (\ref{N_e_1}) with $e$ replaced by HII and HeII, respectively, and $Q$ replaced by their respective collision terms which one can read off from Section 3 of \cite{1999ApJ...523L...1S}.

\subsection{Kinetic matter temperature perturbations}

In this subsection we derive the equation for the perturbation of the kinetic matter temperature. By matter, we refer to the collection of electrons and H and He ions and atoms in all energy levels. Collisions and Coulomb scattering are very fast, and force all these components to have the same temperature and the same velocity $\vec v_b$.

The matter temperature measures the thermal kinetic energy of the matter species, therefore, we look for an equation which includes all important effects which can change the kinetic energy of the matter species. Apart from the usual momentum redshifting, those include electron Compton scattering, bremsstrahlung, and other subleading interactions. Both Compton scattering and bremsstrahlung keep $T_M$ and $T_R$ in thermal contact, but since bremsstrahlung has a timescale about $10^2$ longer than Compton scattering, we can safely neglect the free-free transitions. 

Another effect comes from the fact that during recombination the actual number of free particles changes. In thermal equilibrium, each particle carries on average an energy equal to $3k_B T_M/2$. When an electron combines with a proton, its kinetic energy is lost to radiation, and now there are fewer particles to share  the remaining energy. This effect is usually referred to as molecular weight change, and it is entangled with another effect, usually called recombination-cooling/phoionization-heating, which takes into account that the kinetic energy of the electron that recombines might actually be higher or lower than the average, which means that the remaining particles will have respectively a lower or higher average kinetic energy, and therefore a lower or higher matter temperature. These effects have been shown in \cite{sasselov} to have a negligible effect in the unperturbed universe. This is true even in the perturbed case, though we keep the molecular weight because the derivation becomes more transparent. Finally, the bound-state $\rightarrow$ bound-state ($bb$) transitions affect the kinetic energy just through a negligible recoil energy,  and we can therefore neglect them.

We can find an equation for the matter temperature by defining a kinetic stress energy tensor for all the matter, $T_{M,\mu\nu}$ and writing down its divergence. The word kinetic stays simply to mean that we are neglecting the differences in potential energy (as we said, this is ok because the $bb$ transitions do not affect the matter kinetic energy). Being a perfect fluid to a very good approximation, we can write the kinetic stress energy tensor in the standard form
\begin{equation}\label{Tmngen}
T_{M}^{\mu\nu}=(\rho_M+\mathrm{p}_M)U^\mu U^\nu+g^{\mu\nu}\mathrm{p}_M\ ,
\end{equation}
where $\rho_M$ and $\mathrm{p}_M$ are the energy density (without takeing into account of the electron potential energy) and pressure of matter measured by a local inertial observer instantaneously at rest with respect to the fluid. They are given by (setting $k_B=1$ from now on):
\begin{eqnarray}\label{TudM}
&&\rho_M = [n_p^{(0)} (m_p+m_e)+n_{He}^{(0)}(m_{He}+2m_e)](1+\delta_b)+\frac{3}{2}\mathrm{p}_M\ , \\ \nonumber
&&\mathrm{p}_M=n_t(T_M^{(0)}+\delta T_M)\ , 
\end{eqnarray}
with
\begin{eqnarray}\label{eq:defdelta}
&&n_n^{(0)}= n_p^{(0)}+n_{He}^{(0)}\ , \nonumber\\
&&n_t^{(0)}= n_n^{(0)}+n_e^{(0)}\ , \nonumber\\
&&n_t= n_n^{(0)}(1+\delta_b)+n_e^{(0)}(1+\delta_e)\ .
\end{eqnarray}
Here $n_p$ counts all protons and H atoms (in all excitations), $n_{He}$ counts all He ions and atoms. In the above we used that the H and He have one and the same overdensity, $\delta_b$, which follows from the first order continuity equation (\ref{num_cons_1}) and the fact that they have a common bulk particle velocity.

In order to write the divergence of this stress-energy tensor, we use the fact that for each of the species
\begin{eqnarray}\label{stressenergy}
T^\mu_{\ \nu}=g_{\mathrm{deg}}\int\frac{d^3P^i}{|P_0|}\sqrt{-g}P^\mu P_\nu f=g_{\mathrm{deg}}\int\frac{d^3 p}{E}P^\mu P_\nu F\ ,
\end{eqnarray}
where in the last passage we have used eq.~(\ref{pi-local}) of App.~\ref{section:A}.
We can write the covariant derivative of $T^{\mu\nu}$ of any species using the distribution function for that species \cite{ehlers}
\begin{eqnarray}
T^{\mu\nu}_{ \ \ ;\nu}=\int\frac{D(P^\mu f)}{d\lambda}\pi\ ,
\end{eqnarray}
where $\lambda$ is the affine parameter defined as in the former subsection. The above equation is a differential geometry identity \cite{ehlers} for the tensor given by (\ref{stressenergy}), i.e. it holds for any species.
By using the geodesic equation $DP^\mu/d\lambda=P^\nu P^\mu_{\ ;\nu}=0$, we can express this in terms of the collision term
\begin{eqnarray}\label{Tcov}
T^{\mu\nu}_{ \ \ ;\nu}=\int\pi\; P^\mu\frac{D f}{d\lambda}=\int \pi\; P^\mu C[f] \ .
\end{eqnarray}
We can apply the above identity also to the sum of all the matter components. In this case, on the right hand side there would be an integral of the electron Compton scattering collision term (the proton Compton scattering is negligible), and of the collision term due to bound-bound and bound-free transitions. As we said, though associated to the emission of even highly energetic photons, these last ones only affect the evolution of the population of the excited states, or equivalently of the potential energy of the bound electrons as it gets diluted into radiation, but they do not affect the matter kinetic energy~\footnote{The recoil energy due to a recombination event is very small.}. Therefore, for what concerns the kinetic part of the stress energy tensor defined above, its divergency is given by simply considering the Compton collision term.
 \begin{eqnarray}\label{1stLAW}
T^{\mu\nu}_{M\ ;\nu}=-\int P^\mu C_{\gamma e}[f_{\gamma}]\pi_{\gamma}\ .
\end{eqnarray}

After neglecting polarization, the second order Compton collision term is given by \cite{dodjub}  by expanding the Klein-Nishina cross section to second order in the baryon velocity (first order in $T_M$). Their expression is further reduced to a useful form in \cite{bartolo} (BMR1 from now on), their eq.~(4.42). In  App.~\ref{section:C}, we give an erratum to the second order collision term given in both papers. Using the corrected expression for the Compton collision term, the $\mu=0$ component of the above integral can be evaluated transforming the momenta to the ones of the local inertial frame at rest with the rest frame of the baryon fluid:
\begin{eqnarray}\label{Kompa}
&&e^0{}_{\hat\nu}\left.\left( \int p^{\hat\nu} C_{\gamma e}[f_{\gamma}]\pi_{\gamma}\right)\right|_{v_b=0 \ \mathrm{frame}}=2\frac{1-\Psi}{a}  \int C_{\gamma e}[F_\gamma]d^3 p_\gamma=-\frac{1-\Psi}{a}\Lambda_C\ ,\nonumber\\
\ee
where
\be
&&\Lambda_C\equiv \frac{4\sigma_T a_{\mathrm{Rad}} k_B}{m_e}n_eT_R^4(T_R-T_M)\ ,
\end{eqnarray}
and $e^\mu{}_{\hat\nu}$ is just the tedrad which gives the coordinate transformation between Newtonian gauge and the local inertial frame at rest with the baryons. To first order, only $e^0{}_0=(1-\Psi)/a$ matters.
Here $\sigma_T$ is the Thomson cross section, and $a_{\mathrm{Rad}}$ is the radiation constant. Note in the above only the monopole radiation temperature $T_R$ enters at first order.
The above collision term is nothing else but the Kompaneets collision term for Compton scattering in a local inertial frame at rest with the baryon fluid. 

Combining the expression for the energy-momentum tensor (\ref{Tmngen}), (\ref{TudM}); the number conservation for each nucleus (\ref{num_cons_1}); the evolution of the electron density (\ref{N_e_0}), (\ref{N_e_1}); and the 0 component of (\ref{1stLAW}) together with (\ref{Kompa}) we obtain an equation for the matter temperature, $T_M$. To zeroth order we obtain
\begin{eqnarray}\label{TM_final}
\dot T_M^{(0)}+2\mathcal{H}T_M^{(0)}+\frac{a Q^{(0)}}{n_t^{(0)}}T_M^{(0)}=\frac{2}{3}\frac{a}{n_t^{(0)}}\Lambda_C^{(0)}\ .
\end{eqnarray}
The term including $Q$ corresponds to the evolution of the molecular weight (the number of particles in the gas changes due to the electron recombination). It has a timescale $\gtrsim 10^{3.5}$ larger than the Kompaneets timescale (shown in Fig. \ref{fig:thermal}) of the term on the right hand side, and has therefore a very small effect. The first order expression for the matter temperature is given by
\begin{eqnarray}\label{deltaTM1}
&&\dot \delta T_M+2\mathcal{H}\delta T_M+\frac{2}{3}T_M\theta_b+\frac{a}{n_t^{(0)}} Q^{(0)}\delta T_M-2T_M^{(0)}\dot\Phi+\nonumber\\
&&+\frac{2}{3}\frac{a}{n_t^{(0)}}\left(\frac{3}{2}T_MQ^{(0)}-\Lambda_C^{(0)}\right)\times\left[\Psi-\frac{n_e^{(0)}\delta_e+n_n^{(0)}\delta_b}{n_t^{(0)}}\right]=\nonumber\\
&&=\frac{2}{3}\frac{a}{n_t^{(0)}}\left(\delta\Lambda_C-\frac{3}{2}T_M\delta Q\right)\ .
\end{eqnarray}
To gain some intuition, notice that the $\dot\Phi$, $\theta_b$ and $\mathcal{H}$ terms can be combined to obtain the local velocity divergence, and that $\Psi$ converts coordinate time to local inertial frame time. Neglecting the evolution of the molecular weight, the above equation for $\delta T_M$ reduces to equation (B.12) in \cite{astro-ph/0702600v2}.

\subsection{Discussion of numerical results}\label{section:discussion}

The evolution of $\delta_e$ and $\delta T_M$ is given by eq.s~(\ref{num_cons_1}), (\ref{N_e_1}), (\ref{delta_K_P}) and (\ref{deltaTM1}), where we also include the contribution from He recombination as explained at the end of sec.~\ref{sec:first_order_eq}. The linear approximation is valid up to the reionization epoch, when the electron density perturbation becomes non-linear.

The cosmological parameters we use are $$(\Omega_b,\,\Omega_\Lambda,\,h,\,T_{\rm cmb}\,,Y_p,\,n_s,\tau)=(0.041,\,0.76,\,0.73,\,2.726,\,0.24,\,0.958,\,0.092)\ .$$ We remind that the peak of the visibility function is at $\eta=288.42\,$Mpc and the present conformal time is $\eta_0=14554.28\,$Mpc. We find it helpful to give the following correspondence 
$$\eta=(70, 288, 500, 1000, 5000)\,{\rm Mpc}\quad 
%{\rm correspond\ to\ redshifts} 
\rightarrow\quad z=(5936, 1090, 507, 174, 9)\ .$$

\begin{figure}[h!]
\begin{center}
%\plotone{f3}
\includegraphics[width=15cm]{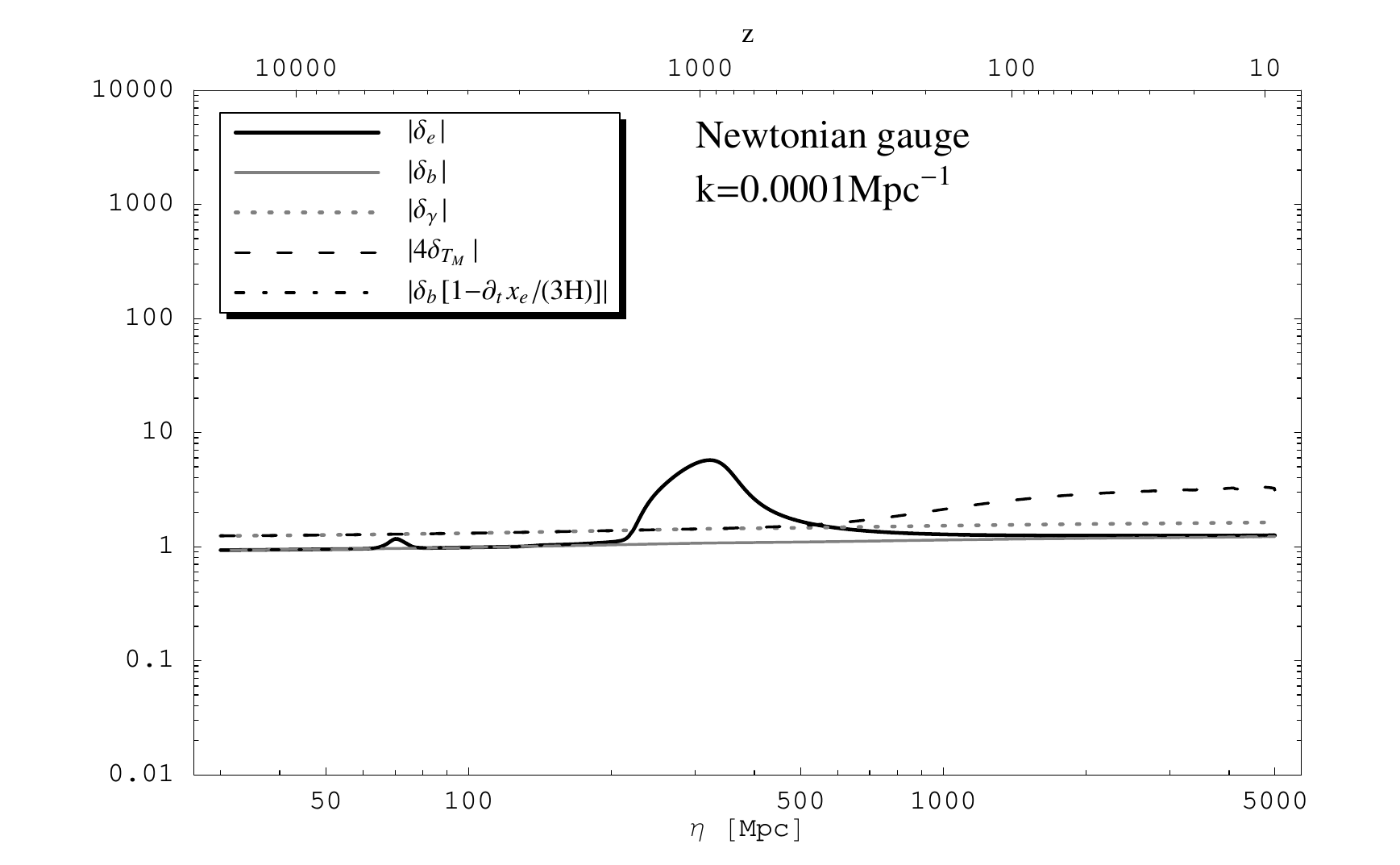}
\end{center}
\caption{The amplitude of the perturbations to the electron density in the Newtonian gauge compared to other first order perturbations for a mode well outside the horizon. The peak of the CMB visibility function is at $\eta=288\,$Mpc. The normalization of the perturbations is the one in CMBFAST -- superhorizon $\zeta=1$. The perturbations in the Newtonian gauge for superhorizon modes are equivalent to time shifted zeroth order evolution. This results in two regions of enhancement of $\delta_e$ corresponding to HeIII$\to$HeII, and the combined HeII$\to$HeI and HII$\to$HI recombination. Similarly, $\delta_{T_M}$ coincides with $\delta_{T_R}$ while Compton scattering is effective in keeping the photons and electrons in thermal contant. As $T_M$ decouples, $\delta_{T_M}$ converges to $2\delta_{T_R}$.} \label{fig:con1}
\end{figure}

In the Introduction we argued that the amplitude of $\delta_e$ at large scales should be enhanced due to the small timescale of recombination. This is valid for gauges, as Newtonian gauge for example, where the perturbations do not vanish for $k\to 0$, and  are completely described as time delayed homogeneous solutions. Therefore, for $k\to 0$ in the Newtonian gauge we expect 
\be\label{dedb}
\delta_e/\delta_b=d \ln n_e/d\ln n_b=1-\partial_\eta (\ln x_e)/(3\mathcal{H})\ ,
\ee
and analogously $\delta_{T_M}/\delta_{T_R}=d \ln T_M/d\ln T_R$. In Fig. \ref{fig:con1} we show a plot of the amplitudes of several perturbations including $\delta_e$ and $\delta_{T_M}$ in the Newtonian gauge for $k=10^{-4}\,$Mpc$^{-1}$. The electron density and matter temperature perturbations indeed follow precisely an evolution equivalent to a time delayed homogeneous universe solution as in eq.~(\ref{dedb}) (in fact in the plot the difference is in practice indistinguishable). The electron density perturbation departs from $\delta_b$ during HeIII$\to$ HeII recombination at $\eta\sim70\,$Mpc, and then during Hydrogen recombination. The HeII$\to$ HeI recombination is delayed \cite{sasselov}, and its contribution to $\delta_e$ overlaps with that of Hydrogen recombination. 

As expected,  the electron density perturbation is enhanced with respect to the other perturbations at recombination.  The plotted first order variables are (except for $\delta_{T_M}$) the matter and radiation fluctuations which source the second order CMB anisotropies. Later we will see that this enhancement persists also for modes well inside the horizon. This  suggests that $\delta_e$ will play an important role in sourcing the bispectrum of the CMB.  The calculation of these effects will be the subject of a companion paper  \cite{inprep}. 

Still in Fig.~\ref{fig:con1}, before $z\sim500$, we notice that $\delta_{T_M}\simeq\delta_{T_R}$ as Compton scattering keeps baryons and photons in thermal contact.  $\delta_{T_M}/\delta_{T_R}$ becomes equal to 2 after $z\sim500$ when the matter temperature decouples. At this points matter and radiation redshift differently even in an unperturbed universe, which explains the factor of 2. 

\begin{figure}[h!]
\begin{center}
%\plotone{f3}
\includegraphics[width=15cm]{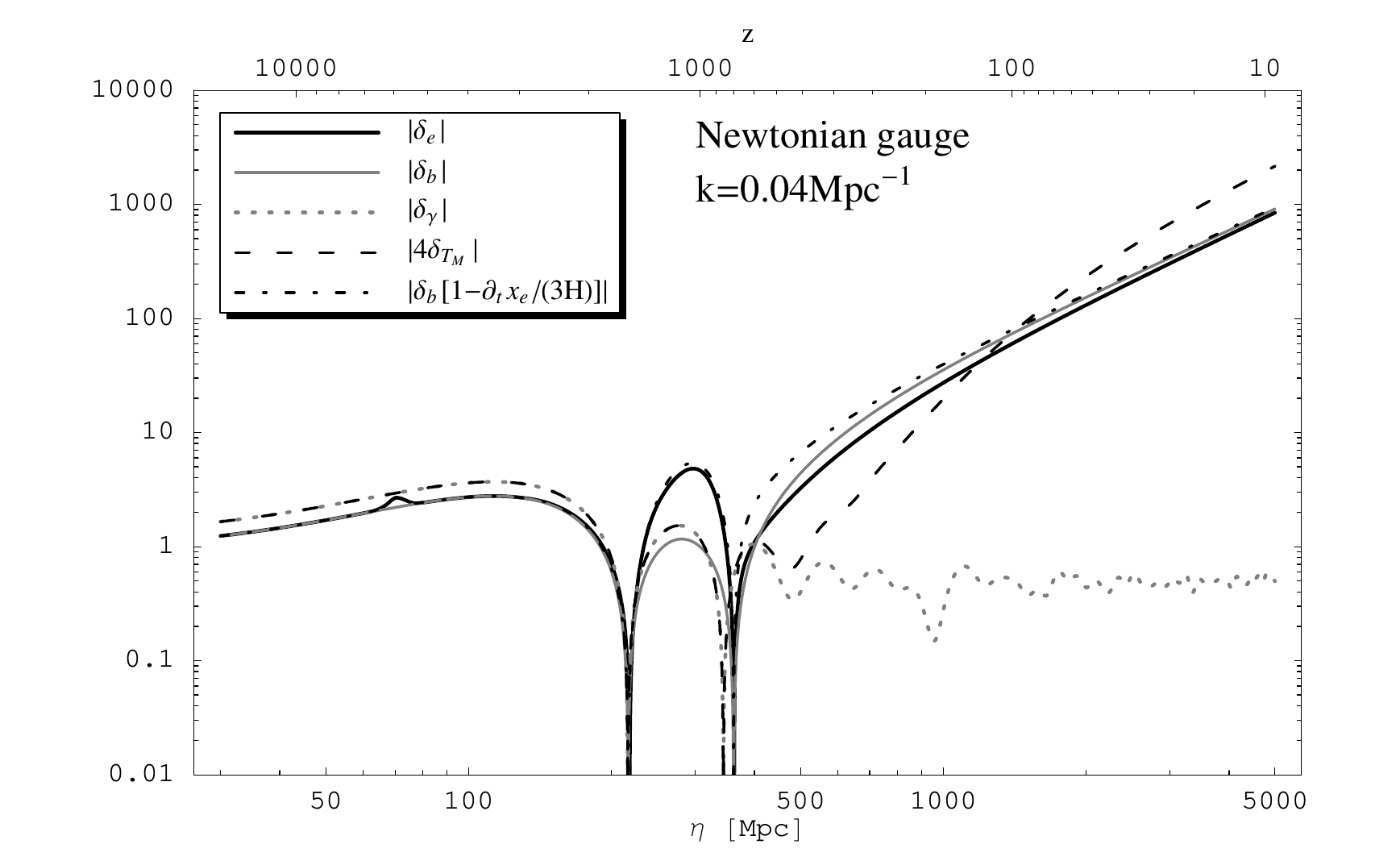}
\end{center}
\caption{The amplitude of the perturbations to the electron density in the Newtonian gauge compared to other first order perturbations for a mode corresponding approximately to the second acoustic peak. The enhancement of $\delta_e$ is still well approximated by eq.~(\ref{dedb}). This mode is well inside the horizon, and therefore, this enhancement is not a gauge artefact, and will leave an imprint on the CMB bispectrum.} \label{fig:con4}
\end{figure}

Even more interestingly, we find that the enhancement to $\delta_e$ persists even for modes inside the horizon where the effect is clearly not gauge dependent. In Fig. \ref{fig:con4} we show $\delta_e$ and $\delta_{T_M}$ for a mode corresponding to the second acoustic peak. The Kompaneets timescale $n_t T_M/\Lambda_C$ is still smaller than one comoving Mpc for $\eta<400\,$Mpc (see Fig. \ref{fig:thermal}), therefore, for the modes we consider, even after recombination, the full (zeroth+first order) $T_M$ and $T_R$ are kept in contact by Compton scattering, as can be seen in Fig. \ref{fig:con4}. At later times, the two temperature fluctuations deviate and $\delta_{T_M}$ becomes greatly enhanced with respect to $\delta_{T_R}$ since overdense regions become hotter due to adiabatic contraction. 

For $z<200$ ($\eta>1000\,$Mpc) one can neglect $\d_\eta \log( x_e)/(3\mathcal{H})\lesssim 0.1$ in (\ref{dedb}), and from the time shift perspective one would expect $\delta_e/\delta_b\approx 1$. However, from Fig. \ref{fig:con4} we clearly see that $\delta_e$ is suppressed compared to $\delta_b$ during the dark ages. As argued in \cite{arXiv:0707.2727v1}, during the dark ages, overdense regions have enhanced $T_M$, and this is associated, as we will soon explain, to a more efficient recombination  and a suppressed $\delta_e$. This effect induces a decrease of about $10\%$ in $\delta_{T_M}$ during the dark ages, which in turn influences the 21$\,$cm power spectrum at a $2\%$ level at $z\sim 50$ \cite{arXiv:0707.2727v1}. For a detailed discussion of the effect of the matter temperature perturbations on the 21cm angular-power spectrum we refer the reader to \cite{astro-ph/0702600v2}.

\begin{figure}[h!]
\begin{center}
%\plotone{f3}
\includegraphics[width=15cm]{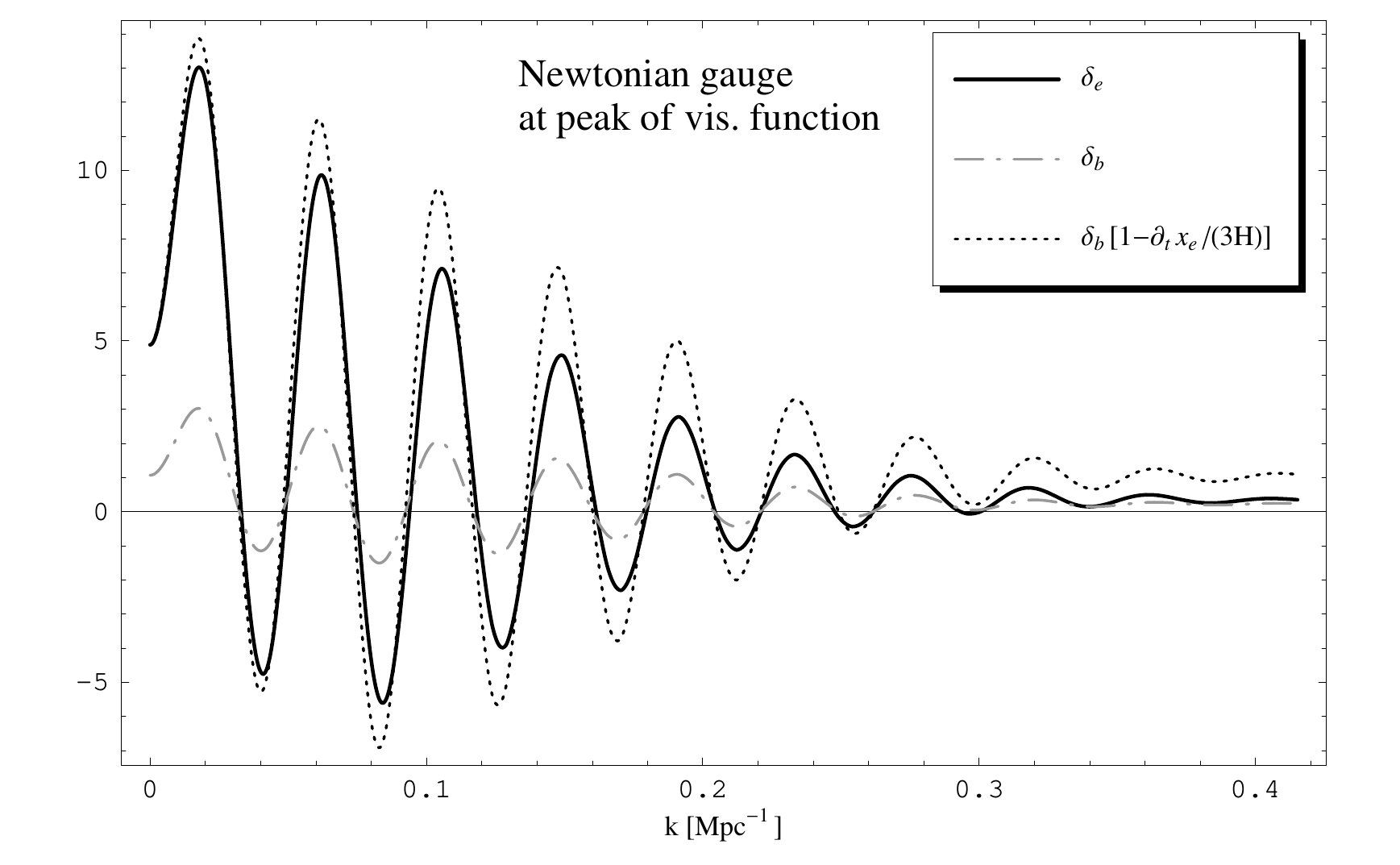}
\end{center}
\caption{A plot of some first order quantities  in the Newtonian gauge at the peak of the visibility function showing the enhancement of $\delta_e$. The first order $\delta_b$ shows the typical magnitude of the variables sourcing the CMB anisotropies at second order, and one can see that $\delta_e$ dominates. We also plot the expected electron density perturbation coming from eq.~(\ref{dedb}) (dotted line) which is a good approximation to the true $\delta_e$ down to the recombination scale ($k\sim0.1\,$Mpc$^{-1}$). The enhancement for superhorizon modes is gauge dependent, but persists for modes inside the horizon, where the effect is observable, approximately down to the photon diffusion scale ($k_D\approx0.15\,$Mpc$^{-1}$).} \label{fig:viscon}
\end{figure}

From the numerical results  (see Figures \ref{fig:con4} and \ref{fig:viscon}), we see that eq.~(\ref{dedb}) is still a good approximation to the ratio of $\delta_e$ and $\delta_b$ even for modes well inside the horizon. This does not mean anymore that $\delta_e$ is just a time delay of the homogeneous solution. Instead, it tells us that $n_e$ is determined by the local value of $n_b$ (or equivalently of the temperature) with the same relationship as in an unperturbed universe: $\delta n_e\simeq \delta n_b\, \dot n_e/\dot n_b$. This can be explained by the fact that the timescale for recombination is still much faster than the mode we are considering.  In Fig. \ref{fig:viscon} we plot $\delta_e$ and $\delta_b$ (which has a size comparable to the other perturbations that source the second order CMB anisotropies) as a function of $k$ at the peak of the visibility function. As one can expect, (\ref{dedb}) well approximates $\delta_e$ down to scales comparable to the recombination timescale ($k\sim0.1\,$Mpc$^{-1}$). The enhancement of $\delta_e$ is prominent for modes with scale larger than approximately $10^{-1}\,$ Mpc$^{-1}$ (which is comparable to the photon diffusion scale $k_D\sim0.15\,$Mpc$^{-1}$). The suppression at high-$k$ with respect to what implied by (\ref{dedb}) is due to the fact that for these modes the timescale of the oscillation becomes faster than the rate of recombination (see Fig.~\ref{fig:recomb}), inducing an averaging out of the enhancement \footnote{This suppression at large $k$ will lead to an important simplification when we derive the CMB bispectrum from recombination \cite{inprep}.}.

Notice that in Fig.~\ref{fig:con4}, we can see that at late times, even for a relatively low-$k$ mode, $|\delta_e|$ is a bit suppressed with respect to what is given by eq.~(\ref{dedb}). Let us briefly explain why this happens. For $\eta\gtrsim 400$ Mpc, the timescale of the mode with $k=0.04$ Mpc$^{-1}$ becomes comparable to the recombination timescale (see again Fig. \ref{fig:recomb}). At this time,  the contribution from Ly$\alpha$ escape, two-photon absorption, and photoionization is negligible, and we can approximate the  homogeneous integrated Hydrogen recombination collision term as   $Q\simeq-\alpha_B n_{\mathrm{HII}}n_e$. This clearly  increases in magnitude for higher $n_e$ or $n_{\rm HII}$.  This dependence of $Q\propto -n_{\mathrm{HII}}$ implies that overdense regions will recombine faster, explaining the suppression of $|\delta_e|$ relative to (\ref{dedb}) at later times in Fig.~\ref{fig:con4}.

%During recombination, the contribution from Ly$\alpha$ escape and two-photon absorption is to a rough approximation negligible (see Fig. \ref{fig:recomb}). Therefore, the homogeneous integrated Hydrogen recombination collision term becomes     $Q \simeq -\alpha_B n_e n_{\mathrm{HII}}\Lambda_{2S\to1S}/(\Lambda_{2S\to1S}+\beta_B)$, which clearly increases in magnitude for higher $n_e$ or $n_{\rm HII}$. This becomes even simpler for $\eta\gtrsim420\,$Mpc, when also the photoionization rate becomes negligible (see again Fig. \ref{fig:recomb}), and thus only recombination remains important. In this regime $Q\simeq-\alpha_B n_{\mathrm{HII}}n_e$. This dependence of $Q\propto -n_{\mathrm{HII}}$ implies that overdense regions will recombine faster, explaining the suppression of $|\delta_e|$ relative to (\ref{dedb}) in Fig.~\ref{fig:con4}.

Let us comment on the validity of our calculation at high $k$'s. In deriving the perturbed Peebles equation, we pointed out that for $k>1\,$Mpc$^{-1}$, the Ly$\alpha$ line cannot be treated as quasi-static, and the perturbations to the escape probability will be modified. However, those perturbations have a comparably small effect on $\delta_e$, since recombination is dominated by two-photon decay. 
The most important effect which may influence the 3-level atom calculation is the effect of nonequilibrium excited levels. That effect in the homogeneous universe leads to a correction of about $10\%$ to $n_e$ for $\eta\gtrsim360$ Mpc, when photoionization becomes negligible (Fig. \ref{fig:recomb}) \cite{sasselov}. At this point recombination to excited states becomes enhanced, and the bound-bound transitions between $n>2$ levels is no more in thermal equilibrium (for a discussion of the effect see \cite{1999ApJ...523L...1S}). This tells us that the timescale over which the excited states equilibrate becomes comparable to the Hubble time  at these redshifts (see again Fig.~\ref{fig:recomb}). Since this effect occurs quite late in the recombination history, it gives only a 10\% effect on $n_e$. In the presence of perturbations with $k$ lower than $\sim 10^{-3}\,$ Mpc$^{-1}$, the effect will be of the same order. For modes with higher $k$, the thermalization timescale becomes more important even earlier, and so potentially the effect could be larger. However, since the photoionization rate grows quite steeply as we move to higher redshifts (see again Fig.~\ref{fig:recomb}), this effect becomes important only for $k\sim 1 \,$ Mpc$^{-1}$. This is the upper limit on the $k$'s for which our calculation is valid.

\subsubsection{Perturbations in the synchronous gauge}

In \cite{inprep} we will be working in synchronous gauge to calculate the CMB bispectrum generated from recombination. In this gauge, the evolution of $\delta_e$ and $\delta T_M$ is given again by (\ref{num_cons_1}), (\ref{N_e_1}), (\ref{delta_K_P}) and (\ref{deltaTM1}), but with the substitutions $\Psi\to 0$, and $\dot \Phi\to -\dot h/6$, where $h$ is the trace of the perturbations to the spatial part of the metric as defined in \cite{maed}. 

\begin{figure}[h!]
\begin{center}
%\plotone{f3}
\includegraphics[width=15cm]{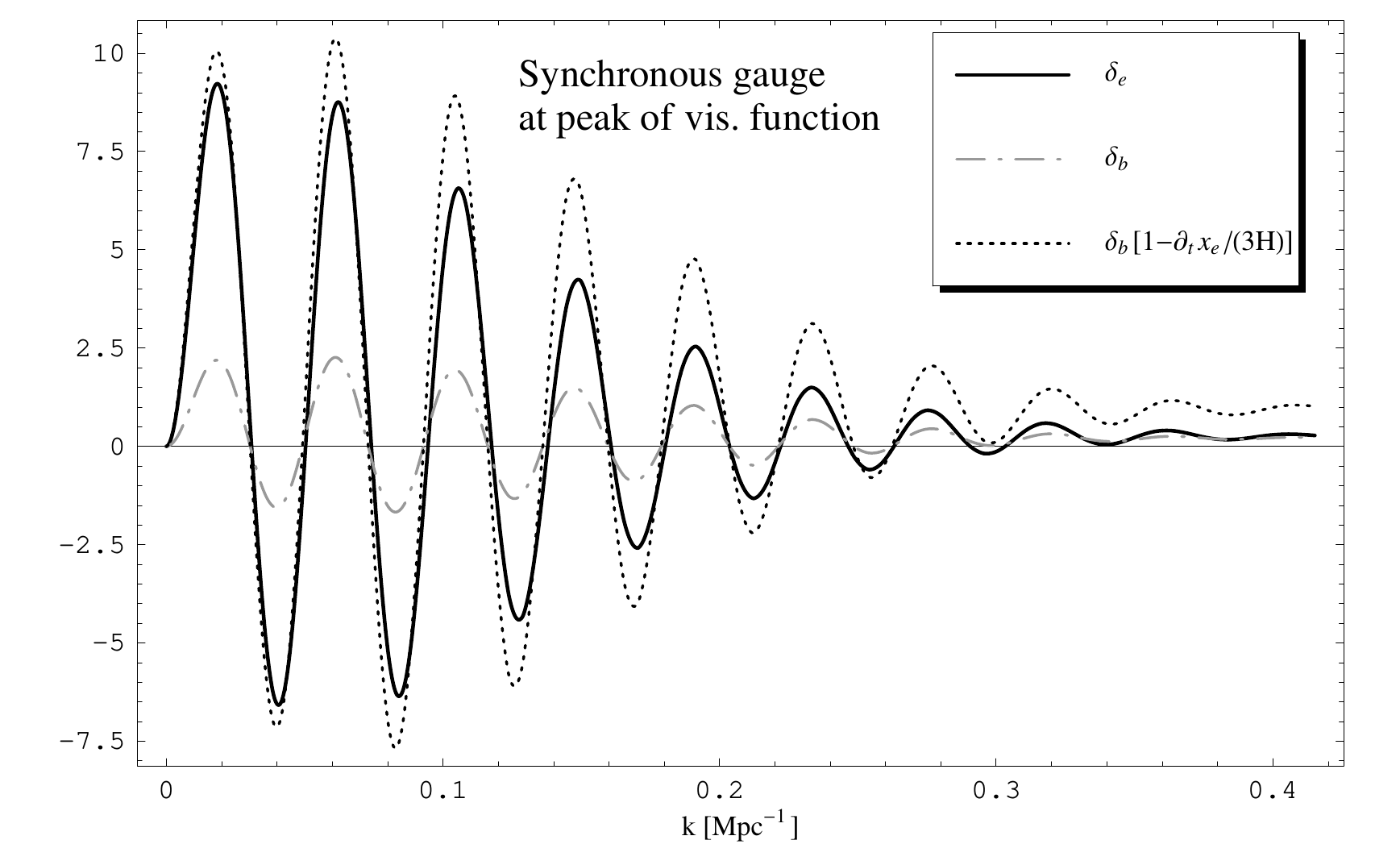}
\end{center}
\caption{The same as in Fig.~\ref{fig:viscon} repeated in the synchronous gauge. The only important difference between the results in the Newtonian and the synchronous gauge is for superhorizon modes for which the amplitude in the synchronous gauge goes to zero.} \label{fig:vissyn}
\end{figure}

In Fig.~\ref{fig:vissyn} we show the results for $\delta_e$ in the synchronous gauge. As expected, for modes well inside the horizon, the density perturbations in the synchronous and Newtonian gauges are nearly identical (cf. Fig.~\ref{fig:viscon}).  Perturbations vanish for superhorizon modes since they correspond to time-shifted homogeneous solutions and the synchronous gauge time coordinate equals the proper time of comoving observers  which makes the time-delay vanish outside of the horizon. As we will show in \cite{inprep}, this property of the synchronous gauge will be very useful in estimating the CMB bispectrum generated from recombination.

\section{Perturbations to Second Order}\label{section:secondorder}

We now begin the second part of this paper, where we will give the full set of second order Boltzmann and Einstein equations that are necessary to compute the non-Gaussian signal introduced in the CMB by the standard cosmological evolution. In the first part, we computed the first order perturbation for $\delta_e$ because, as we will see, it will be a source for the second order perturbations. In this derivation, we will follow quite closely BMR1 (i.e. Ref.~\cite{bartolo}), who first gave the full set of second order equations, though in our derivation there will be a few important differences.

\subsection{The Poisson gauge}
The metric in Poisson gauge \cite{edbert} is given by 
\begin{eqnarray}\label{metric}
ds^2=a^2(\eta )\left[-e^{2\Psi}d\eta ^2+2\omega_i dx^i d\eta +(e^{-2\Phi}\delta_{ij}+\chi_{ij})dx^idx^j\right]\ ,
\end{eqnarray}
with $\chi_{ii}=0$, since the trace can always be absorbed in $\Phi$.
The gauge fixing is done by choosing 
\begin{eqnarray}\label{gauge_fix}
\omega_{i,i}=0\ , \ \ \ \ \ \chi_{ij,j}=0\ ,
\end{eqnarray}
for each order in perturbation theory \cite{bruni}. Following \cite{bartolo} we restrict the Poisson gauge by choosing the vector and tensor modes given by $\omega_{i}$ and $\chi_{ij}$ to be second order quantities  (we neglect here primordial vector and tensor modes). For convenience in the expressions that follow, we will often use $e^{2\Psi}$ and $e^{2\Phi}$. In this gauge, there is one scalar degree of freedom eliminated from $g_{0i}$ and one scalar and one transverse vector degrees of freedom from $g_{ij}$. Thus, $\omega_i$ is a transverse vector, while $\chi_{ij}$ is a traceless-transverse tensor. The scalar potentials $\Psi$ and $\Phi$ contain the first and second order potentials as follows: $\Psi=\Psi^{(1)}+\Psi^{(2)}/2$, $\Phi=\Phi^{(1)}+\Phi^{(2)}/2$, where $\Phi^{(2)}$ and $\Psi^{(2)}$ are meant to be of order $(\Phi^{(1)})^2$. We notice also the nice property that $\sqrt{-g}=e^{\Psi-3\Phi}a^4$ to second order. Notice that, with respect to BMR1's conventions, we substituted $\Phi \leftrightarrow\Psi$, so that they match the usual notation in the Newtonian conformal gauge. We use the time-independent piece of the spatial part of the homogeneous FRW metric, $\delta^{ij}$ and $\delta_{ij}$, to raise and lower spatial indices. Thus, we do not keep track of the placement of spatial indices \footnote{Since in this section we are summarizing second order results  in Poisson gauge, we find it useful to mention that BMR give the correct second order Christoffel symbols for this gauge in eq.~(A.2) in their \cite{astro-ph/0703496} (BMR1 have instead a sign typo in their eq.~(2.8) for the Christoffel symbols). }.

In order to derive the Boltzmann equation it is useful to find the change of coordinate from Poisson gauge to a local inertial frame. This can be found by solving for the tetrad $e^\mu_{\ \alpha'}$ which satisfies
\be
e^\mu_{\ \alpha'}e^\nu_{\ \beta'}g_{\mu\nu}=\eta_{\alpha'\beta'}\ .
\ee
%From (\ref{tetrad_P}) we can find the transformation between the conjugate momentum $P_\mu$ and the inertial momentum $p^i=p_i$ measured by comoving observers. Solving the matrix equation for the tetrads, $e^\mu_{\ \alpha'}e^\nu_{\ \beta'}g_{\mu\nu}=\eta_{\alpha'\beta'}$, to second order we get for $e^\mu_{\ \alpha'}$ (dropping the primes for clarity):
Solving to second order we get:
\begin{eqnarray}\label{tetrad2}
e^0_{\ 0}=\frac{e^{-\Psi}}{a}\ , \ \ \ \ \ e^0_{\ i}=\frac{\omega_i}{a}\ , \ \ \ \ \ e^i_{\ 0}=0\ , \ \ \ \ \ e^i_{\ j}=\frac{1}{a}\left[e^{\Phi}\delta_{ij}-\frac{1}{2}\chi_{ij}\right]\ .
\end{eqnarray}
This tetrad is not unique since the local observer coordinates can be Lorentz transformed. The momenta in the two frames are related by the tetrad in the following way:
\begin{equation}
P^\mu=e^\mu_{\ \nu'}p^{\nu'}\ ,
\end{equation}
where $P^\mu$ are the momenta in the Poisson gauge, and $p^{\mu'}$ are the ones in the local inertial frame. The tetrad above is chosen such that $P^i=0$ for particles at rest with respect to the inertial observer (i.e. when $\vec{ p}=0$), which allows for an easy interpretation of the observers' trajectories (see below).
Using (\ref{tetrad2}) and writing $p_ip^i\equiv p^2$ and $p^i\equiv p\, n^i$ (such that $n^in_i=1$), we obtain the following relation between Poisson gauge momenta and local inertial frame momenta~\footnote{Note that the expression eq.~(3.6) for $P^i$ in \cite{bartolo} is not a linear function of $p^\mu$. Thus, their $p^\mu$'s that are present in the streaming terms are {\it not} momenta measured in a local inertial frame. This leads to an inconsistency since the $p^\mu$ appearing in their collision term should be the momenta measured in a local inertial frame. Thus, the streaming terms derived in the next section are different from the ones in BMR1. However, it turns out that this has no effect on the streaming term of the final equations for the evolution of the baryon and CDM number density, on their continuity equations, as well as on the photon brightness equation at second order, because of what seems to us to be just accidental cancellations.}:
\begin{eqnarray}\label{P}
&&P^0=\frac{E}{a(\eta )}e^{-\Psi}\left[1+\frac{p}{E}\omega_in^i\right]\ , \ \ \ \ \ 
P^i=\frac{p}{a(\eta )}e^{\Phi}n^j\left(\delta_{ij}-\frac{1}{2}\chi_{ij}\right)\nonumber\ , \\
&&P_0=-a E e^{\Psi}\ , \ \ \ \ \ P_i=a e^{-\Phi}\left(p_i+E \omega_i+\frac{1}{2}\chi_{ij}p^j\right)\ .
\end{eqnarray}

The observer's trajectory is given by $dx^i/d\eta |_{\vec{ p}=0}=P^i/P^0|_{\vec{ p}=0}=0$. Thus, the local  inertial frame that we have defined in this way is nothing but the Local Inertial Frame Instantaneously at Rest with respect to Comoving Observer (LIFIRCO).

\subsection{Streaming terms}

We are now ready to begin to write down the Boltzmann equation. From (\ref{boltz1}) of App.~\ref{section:A}, we can write the Boltzmann equation for the diffeomorphism invariant one-particle distribution function, $f$, as~\footnote{In order to help comparison with the existing literature, we notice that  this equation is equivalent to eq.~(4.1) in \cite{astro-ph/0703496}, but differs from eq.~(3.1) in \cite{bartolo}. }
\begin{eqnarray}\label{boltz}
P^0\frac{d\, f}{d\eta }=P^\mu\frac{\d\, f}{\d x^\mu}+\frac{d P^i}{d\lambda}\frac{\d\, f}{\d P^i}=C[f]\ ,
\end{eqnarray}
where $C[f]$ is the diffeomorphism invariant collision functional. As we explain in App.~\ref{section:A} and as we have already used in the former section, it is useful to introduce 'nice' coordinates where the momenta are the ones of the local inertial frame, and we define 
\be
F(x^\mu,p^j)\equiv f(x^\mu,P^j(p^i,x^\mu) )\ .
\ee
In this way the collision term is the same as in Minkowski space, and all the metric perturbations are confined to the free-streaming term. In these nice coordinates, the Boltzmann equation becomes
\begin{eqnarray}\label{boltz2}
\frac{\partial F}{\partial \eta }+\frac{\partial F}{\partial x^i}\frac{dx^i}{d\eta }+\frac{\partial F}{\partial p^i}\frac{dp^i}{d\eta }=\frac{1}{P^0}C[F]\ ,
\end{eqnarray}
or equivalently~\footnote{For simplicity, we denote both $F(x^i,p^j; \eta )$ and $F(x^i,p,n^j; \eta )$ with the same symbol $F$. It is clear which one is used in all equations below.} 
\begin{eqnarray}\label{boltz3}
\frac{\partial F}{\partial \eta }+\frac{\partial F}{\partial x^i}\frac{dx^i}{d\eta }+\frac{\partial F}{\partial p}\frac{dp}{d\eta }+\frac{\partial F}{\partial n^i}\frac{dn^i}{d\eta }=\frac{1}{P^0}C[F]\ .
\end{eqnarray}

To derive the Boltzmann equation to second order, we need to calculate $dp/d\eta $, $dp^i/d\eta $ and $dx^i/d\eta $ to second order, and $dn^i/d\eta $ to first order, since $\partial F/\partial n^i$ is already a first-order quantity. 

Writing the coordinate velocity $dx^i/d\eta$ to second order is simple: 
\begin{eqnarray}\label{x}
\frac{dx^i}{d\eta }=\frac{P^i}{P^0}=\frac{p}{E}n^j e^{\Phi+\Psi}\left[\delta_{ij}\left(1-\frac{p}{E}\omega_kn^k\right)-\frac{1}{2}\chi_{ij}\right]\ .
\end{eqnarray}
The $\omega$ piece in (\ref{x}) converts $E$ as observed by LIFIRCO observers  to the energy of the particle as measured in the normal inertial frame, defined as the frame which is free falling,  which moves with a velocity ($-\omega^i$) with respect to the LIFIRCO observers \cite{edbert}. The rest of the terms appearing above convert from inertial frame velocity to coordinate velocity.

To obtain $dp^i/d\eta $ and $dn^i/d\eta $ we write the geodesic equation
\begin{eqnarray}\label{geo_i}
\frac{dP^\mu}{d\eta }=-\Gamma^\mu_{\nu\rho}\frac{P^\nu P^\rho }{P^0}\ ,
\end{eqnarray}
which, by using  (\ref{P}), and after some simple manipulations, give:
\begin{eqnarray}\label{eq:ni}
&&\frac{dn^i}{d\eta }=-\left(\delta^{ij}-n^in^j\right)\left(\frac{E}{p}\Psi_{,j}+\frac{p}{E}\Phi_{,j}\right)\ , \\ %\\
%n^in^k\left[\frac{E}{p}\Psi_{,k}+\frac{p}{E}\Phi_{,k}\right]-\frac{E}{p}\Psi_{,i}-\frac{p}{E}\Phi_{,i}
\label{dpdt}
&&\frac{1}{p}\frac{dp}{d\eta }=-\mathcal{H}+\dot\Phi -\frac{E}{p}n^i\Psi_{,i}e^{\Phi+\Psi}-\frac{1}{2}n^in^j\dot\chi_{ij}-\frac{E}{p}n^i\dot\omega_i-\frac{m^2}{Ep}\mathcal{H}\omega_in^i\ .
\end{eqnarray}
As one could have expected, $dn^i/d\eta$ is affected only by the transverse gradients of the scalar potentials. The second term in (\ref{eq:ni}), $\Phi_{,j}$, which for nonrelativistic particles is $\mathcal{O}(v^2)$ suppressed compared to the $\Psi_{,j}$ term, is responsible for the fact that relativistic particles are deflected twice as much compared to nonrelativistic particles.

We can also write $dp^k/d\eta $ to second order
\begin{eqnarray}\label{ppp}
&&\frac{1}{p}\frac{dp^k}{d\eta }=-n^k\mathcal{H}+n^k\dot\Phi-e^{\Phi+\Psi}\left[\frac{E}{p}\Psi_{,k}+\frac{p}{E}\Phi_{,k}-\frac{p}{E}n^kn^i\Phi_{,i}\right]-\frac{m^2}{pE}\mathcal{H}\omega_k+\nonumber\\
&&+n^i\left(\omega_{i,k}-\omega_{k,i}\right)-\frac{E}{p}\dot\omega_k-\frac{1}{2}n^i\dot\chi_{ik}-\frac{p}{2E}n^in^j\left(\chi_{ik,j}-\chi_{ij,k}\right)\ ,
\end{eqnarray}
where a dot denotes $\partial/\partial \eta $, and $\mathcal{H}\equiv \dot a/a$~\footnote{For the sake of comparison, we can rewrite the above equation in a form similar to equation (4.70) of \cite{edbert}:
\be\label{ppp1}
&& e^{-\Psi}\frac{d}{d\eta}\left[a e^{-\Phi}\left(1+\frac{1}{2}\chi\cdot\right)\vec{p}\right]=\nonumber\\
&& a\, E \left[-\vec{\nabla} \Psi-\dot{\vec{\omega}}+\frac{1}{E}\vec{p}\times (\vec{\nabla}\times\vec{\omega})-\left(\frac{p}{E}\right)^2\vec{\nabla}\Phi+\frac{1}{2}\left(\frac{p}{E}\right)^2n^in^j\vec{\nabla} \chi_{ij}\right]-\partial_\eta(a E) \,\vec{\omega}
\ee
where we defined $[\chi\cdot\vec{p}]_i\equiv \chi_{ij}p^j$. The above equation is correct to second order in the metric perturbations. The reader is referred to equation (4.70) of \cite{edbert} and its discussion for the interpretation of the terms appearing above. Since here $\omega_i$ and $\chi_{ij}$ are second order, and $\Psi$ and $\Phi$ contain the second order scalar perturbations, the only additional component to (4.70) of \cite{edbert} is the extra factor of $e^{-\Psi}$ in (\ref{ppp1}) which converts coordinate time to proper time.}.

\subsection{Evolution of the number density and momentum conservation at second order}

%From now on, with $n_j$ we will denote the number density of the $j^\mathrm{th}$ particle species as measured by a local comoving observer at fixed $x^i$, unless stated otherwise. If there is a chance for confusion between the number densities and the components of $\hat{n}$, the difference will be noted. The proper number density, $n_j$, of the $j^\mathrm{th}$ patricle species is the density measured by an observer at rest with respect to the fluid.

The particle 4-current measured in the LIFIRCO, which we call $N^\mu$, can be written using eq.~(\ref{current}) of App.~\ref{section:A}:
\be\label{num}
N^\mu\equiv g_{\mathrm{deg}}\int \frac{d^3p}{E} f(\vec{x},P_j(\vec{ x},\vec{ p};\eta );\eta )p^\mu\ .
\ee
The average fluid velocity $v^i$ measured in the LIFIRCO is given by
\begin{equation}\label{vel}
v^i=\frac{N^i}{N^0}\ .
\end{equation}
It is good to know that the transformation from the LIFIRCO to the local inertial frame momentarily at rest with the fluid is given by a boost with velocity  $v^i$~\footnote{Under a boost with velocity $\vec{\mathrm{v}}$ we can go from the LIFIRCO to the instantaneous rest frame of the fluid. Let us check that the average velocity $\vec v$ of the fluid coincides with $\vec{\mathrm{v}}$. We can use the invariance of $d^3p/E$ (\ref{pi}) and $f$ to rewrite (\ref{vel}) as 
\begin{equation}\label{vel1}
v^i=g_{\mathrm{deg}}\frac{1}{N^0}\int \frac{d^3p'}{E'} f(\vec{ x},P_j(\vec{ x},\vec{ p}';\eta );\eta )p^i\ .
\end{equation}
Using the transformation of $p^\mu$ under a boost with velocity $\vec{\mathrm{v}}$ such that we go to the rest frame of the fluid, we can express $\vec{p}$ in terms of $\vec{p}'$ and $E'$, with which we denote the momentum and energy of the particles measured in the fluid rest frame. Assuming that the distribution function has vanishing dipole in the rest frame of the fluid (by definition of the fluid rest frame), from (\ref{num}) evaluated in the fluid rest frame and (\ref{vel1}) we obtain
\begin{equation}\label{vel2}
\vec{v}=\frac{n}{N^0}\frac{\vec{\mathrm{v}}}{\sqrt{1-\vec{\mathrm{v}}^2}}=\vec{\mathrm{v}}\ .
\end{equation}
Thus, $\vec{\mathrm{v}}$ coincides with the average velocity of the fluid $\vec v$.}. This implies the standard relation $N^i=N^0v^i=n\, v^i/\sqrt{1-v^2}$. To get an equation for $N^0$, we can follow BMR1 and multiply both sides of (\ref{boltz2}) by $P^0/E$ and integrate over $d^3p$. Another approach which simplifies the calculation is to use (\ref{boltzmannequation}) and (\ref{coll}) of App.~\ref{section:A} to write the conservation of the particle 4-current as ${\cal{N}}^\mu_{\ ;\mu}=\int\pi\, C[f]$ which, by expressing ${\cal{N}}^\mu$ in Poisson gauge in terms of $N^\mu$ by the tetrad relation (\ref{tetrad2}), to second order gives: 
\be\label{num_evol}
&&\frac{1}{a} e^{-\Psi}\left[\dot N^0+3N^0(\mathcal{H}-\dot\Phi)+e^{\Phi+\Psi}N^i_{,i}+e^{\Phi+\Psi}N^i(\Psi_{,i}-2\Phi_{,i})\right]\nonumber\\
&&= g_{\mathrm{deg}}\int \frac{d^3p}{E} C[F]\ .
\ee
In this equation all the variables are meant to be the full variables, containing their zeroth, first and second order pieces~\footnote{This equation is also given by BMR1, but they set the right hand side to zero even for $n_e$, which is an approximation that during recombination is not satisfied. }. Notice that for the perturbation $\delta_b$ to the nuclei number as defined in (\ref{eq:defdelta})  the right hand side above vanishes. 

Next we derive the momentum conservation equation for baryons (meant as usual as the collection of ions, electrons, and atoms). In doing that, one has to be careful in treating the momentum transfers in radiative transitions that do not affect relevantly the kinetic energy of the fluid, and in the  Coulomb interactions, which force electrons and ions to have the same velocity. We already discussed these effects when we derived the perturbations to the kinetic matter temperature. Thus, we derive momentum conservation from the $i$th component of (\ref{1stLAW}) which involves only the photon second order Compton collision term (given in our eq.~(\ref{compton2}) in App.~\ref{section:C}). Unless stated otherwise, we denote the order in perturbation theory of all variables by a superscript in parenthesis, which we drop whenever this does not cause any confusion. Also, for simplicity, we drop  the subscript $_b$ from the baryon velocity $\vec v_b$. Thus, the first and second order expressions for the baryon momentum are given by:
\be
\frac{\partial v^{(1)i}}{\partial\eta}
+\mathcal{H}v^{(1)i}+\Psi^{(1)}_{,i}=\frac{4}{3}\dot\tau\frac{\rho_\gamma^{(0)}}{\rho_b^{(0)}}\left(v^{(1)i}-v^{(1)i}_\gamma\right)\ ,
\ee
which coincides with BMR1, and~\footnote{Note that the left hand side of this expression is the same as eq.~(6.44) in \cite{bartolo} (with the substitutions $\Psi\to\Phi$, $\Phi\to\Psi$). However, the right hand side has acquired a dependence on $\delta_e$.}
\begin{eqnarray}\label{v2}
%\frac{1}{2}\left(\frac{\partial v^{(2)i}}{\partial \eta}+\mathcal{H}v^{(2)i}+2\dot\omega^i+2\mathcal{H}\omega^i+\Psi^{(2)}_{,i}\right)-\dot\Phi^{(1)}v^{(1)i}+v^{(1)j}
%v^{(1)i}_{\ \ \ ,j}+(\Psi^{(1)}+\Phi^{(1)})\Psi^{(1)}_{,i}\nonumber\\
%=\frac{4}{3}\dot\tau\frac{\rho_\gamma^{(0)}}{\rho_b^{(0)}}\left[\left(\Delta_0^{(1)}+\Psi+\delta_e-\delta_b\right)(v^{(1)i}-v^{(1)i}_\gamma)+\left(\frac{v^{(2)i}}{2}-\frac{v^{(2)i}_\gamma}{2}\right)+\frac{3}{4}v^{(1)j}\Pi^{ji}_\gamma\right]
&&\frac{1}{2}\left(\frac{\partial v^{(2)i}}{\partial \eta}+\mathcal{H}v^{(2)i}+2\dot\omega^i+2\mathcal{H}\omega^i+\Psi^{(2)}_{,i}\right)-\dot\Phi^{(1)}v^{(1)i}+v^{(1)j}
v^{(1)i}_{ \ \ ,j}+\nonumber\\
&&+(\Psi^{(1)}+\Phi^{(1)})\Psi^{(1)}_{,i}=\frac{4}{3}\dot\tau\frac{\rho_\gamma^{(0)}}{\rho_M^{(0)}}\left[\left(\Delta_0^{(1)}+\Psi+\delta_e-\delta_b\right)(v^{(1)i}-v^{(1)i}_\gamma)+\right.\nonumber\\
&&\left.\left(\frac{v^{(2)i}}{2}-\frac{v^{(2)i}_\gamma}{2}\right)+\frac{3}{4}v^{(1)j}\Pi^{ji}_\gamma\right] \ .
\end{eqnarray}
In analogy to BMR1,  we have found it useful to define the photon velocity as
\be
v^i_\gamma=\frac{1}{\rho_\gamma+p_\gamma}\int d^3p\, F_\gamma\, p^i \ .
\ee
$\dot\tau=-n_e^{(0)}\sigma_T a$ is the differential optical depth, $\Delta_{0}^{(1)}=\delta_\gamma$, and $\Pi^{ji}_\gamma$ is the photon quadrupole defined as
\be
\label{eq:photon_quadr}
\Pi_{\gamma}{}^i{}_{ j} = \int \frac{d \Omega}{4 \pi} \left( n^i n^j - \frac{1}{3}
\delta^{ij} \right) \left( \Delta^{(1)} + \frac{\Delta^{(2)}}{2}
\right)\ .
\ee
We also neglect terms very small suppressed by $T_M/m_p$, where $m_p$ is the proton mass. The term proportional to $\Delta_{0}^{(1)}+\delta_e-\delta_b$ on the right hand side is the perturbation to $\dot\tau\rho^{(0)}_\gamma/\rho^{(0)}_b$, and as usual $\Psi$ converts coordinate to proper time. The velocity measured by comoving  LIFIRCO observers appearing on the left hand side is converted to the velocity in the normal frame, $\vec{v}+\vec{\omega}$. The combination $\dot {\vec{v}}+v^i\vec{v}_{,i}$ converts the Eulerian time derivative to Lagrangian derivative. The second order piece of $e^{\Psi+\Phi}\Psi_{,i}$ corresponds to the contribution to the acceleration from the gradient of the Newtonian potential with the appropriate conversions between inertial and comoving coordinates. The $\dot\Phi$ term accounts for the usual redshifting of momentum in a time-varying potential.

\subsection{Photon brightness equation}

From the Boltzmann equation for the photons, we get the second order brightness equation
%\begin{scriptsize}
\be
&&\frac{1}{2}\frac{d}{d\eta}\left[\Delta^{(2)}+4\Psi^{(2)}\right]+\left.\frac{d}{d\eta}\left[\Delta^{(1)}+4\Psi^{(1)}\right]\right\vert^{(2)}-4\Delta^{(1)}(\dot\Phi^{(1)}-\Psi^{(1)}_{,i}n^i)\\ \nonumber 
&&-2\frac{\partial }{\partial \eta}(\Psi^{(2)}+\Phi^{(2)})+4\dot \omega_in^i+2\dot\chi_{ij}n^in^j
\\ \nonumber 
&&=-\frac{\dot \tau}{2}\bigg[
\Delta^{(2)}_{00}-\Delta^{(2)}
-\frac{1}{2}\sum_{m} \sqrt{\frac{4\pi}{5^3}}\Delta^{(2)}_{2m}Y_{2m}(\hat {n})+
4 \vec{n} \cdot \vec{v}^{(2)}\\ \nonumber
&&+2(\delta_e+\Psi)\left(\Delta_{00}^{(1)}-\Delta^{(1)}-\frac{1}{2}\sum_{m} \sqrt{\frac{4\pi}{5^3}}\Delta^{(1)}_{2m}Y_{2m}(\vec{ n})+4 \vec{v}^{(1)} \cdot \vec{n}\right) 
\\ \nonumber 
&&+14(\vec v\cdot\hat n)^2-2v^2
+2\vec v\cdot \hat n\left[\Delta^{(1)}+3\Delta^{(1)}_{00}-\frac{1}{2}\sqrt{\frac{4\pi}{5}}\sum_m\Delta_{2m}^{(1)}Y_{2m}(\hat n)+i\sqrt{\frac{\pi}{3}}\sum_m \Delta_{1m}^{(1)}Y_{1m}(\hat n)\right]\\ &&\nonumber  
+
4\pi v\sqrt{\frac{2}{15}}\sum_{m,M}\left(
\begin{array}{ccc}
1 & 1 & 2 \\
m & M & -m-M
\end{array}
\right)\Delta_{2,m+M}^{(1)}Y_{1m}(\hat n)Y_{1M}(\hat v)(-1)^{m+M}\\ \nonumber
&&+2 i\sqrt{\frac{\pi}{3}}v\sum_m Y_{1m}(\hat v) \Delta_{1m}^{(1)}
\bigg]\ ,
\ee
%\end{scriptsize}
where we have used the Compton collision term given in eq.~(\ref{compton2}) and we have explicitly reminded that in the second term in the first line one has to take just the second order contribution. Here we have defined 
\be
\Delta^{(i)}(\vec x,\eta,\hat n)=\frac{\int dp\, p^3 F^{(i)}}{\int dp\, p^3 F^{(0)}}\ ,
\ee
where we have expanded the distribution function as
\be
F(\vec x,p^i,\eta)=F^{(0)}(\vec x,p^i,\eta)+F^{(1)}(\vec x,p^i,\eta)+\frac{1}{2}F^{(2)}(\vec x,p^i,\eta)\ ,
\ee
where $F^{(1)}$ is the first order perturbation and $F^{(2)}$ is the second order one. The $p^i$'s are the the moments as measured in the LIFIRCO. The expansion in sperical harmonics of a quantity $\Delta(\hat n)$ is given by:
\be
\Delta_{lm}=(-i)^{-l}\sqrt{\frac{2l+1}{4\pi}}\int d\Omega \Delta(\hat n)Y^\star_{lm}(\hat n)\ .
\ee

\subsection{Energy-Momentum tensor}\label{section:ein}
We now give an expression for the second order energy momentum tensor and for the Einstein equations in the remainder of this section. In what follows, primed quantities are meant to be evaluated in the particular inertial frame that is at rest with the fluid. The $00$ component of the energy-momentum tensor given in (\ref{stressenergy}) can be written for the matter components using (\ref{P}) in the form
\begin{equation}\label{T2}
T^{\, 0}{}_{ 0}=-g_{\mathrm{deg}}\int d^3 pE F=-g_{\mathrm{deg}}\int \frac{d^3 p'}{E'}E^2 F'=-g_{\mathrm{deg}}\int \frac{d^3 p'}{E'}\frac{E'^2+(\vec{v}\cdot \vec{ p}')^2}{1-v^2} F'=-\frac{\rho+v^2 \mathrm{p}}{1-v^2}\ ,
\end{equation}
where $\rho$ and $\mathrm{p}$ are the energy density and the pressure of the fluid. For the photons we have
\be
T_\gamma^{\,0}{}_{\, 0}=-2 \int d^3 pE F_\gamma=-\rho_\gamma^{(0)}\left(1+\Delta_0^{(1)}+\frac{1}{2}\Delta_{00}^{(2)}\right)\ .
\ee
The other components of $T^\mu{}_{\nu}$ are given by
\be
&&T^i_{\ 0}=-g_{\mathrm{deg}}e^{\Psi+\Phi}\int d^3\, p F\, p^i\ , \\
&&T^0_{\ i}=g_{\mathrm{deg}}e^{-\Psi-\Phi}\int d^3 p\, F \left[p^i +\omega^i\left(E+\frac{1}{3}\frac{p^2}{E}\right)\right]\ , \\ \label{eq:Tij}
&&T^i_{\ j}=g_{\mathrm{deg}}\int \frac{d^3 p}{E} Fp^ip_j\ .
\end{eqnarray}
Let us elaborate on these expressions. From the last equation above we can extract the expression for the pressure $\mathrm{p}$:
\begin{equation}\label{press}
\mathrm{p}=\left.\frac{1}{3}T^i_{\ i}\right|_{v_b=0\ \mathrm{frame}}=\frac{1}{3}g_{\mathrm{deg}}\int \frac{d^3 p'}{E'}p'^2 F'\ ,
\end{equation}
confirming what we used in (\ref{T2}).
For massless particles, this automatically implies the usual $\rho=3\mathrm{p}$. Using (\ref{press}), we can write to second order
\begin{eqnarray}\label{Toi}
T^0_{\ i}=g_{\mathrm{deg}}e^{-\Psi-\Phi}\int d^3 p F p^i+\omega^i(\bar\rho+\bar{\mathrm{p}})=-e^{-2(\Psi+\Phi)}T^i_{\ 0}+\omega^i(\rho+\mathrm{p})\ .
\end{eqnarray}
For non-relativistic particles, the above integrals can be evaluated in a variety of ways. One can work with the distribution function in the rest frame of the fluid, following the steps we used to derive $T^0_{\ 0}$ in (\ref{T2}). Or one can expand the relativistic Boltzmann distribution function in the comoving frame. Yet another way is to write the energy-momentum tensor in the standard covariant form
\begin{equation}\label{Tmncov}
T^{\mu}_{\ \nu}=(\rho+\mathrm{p})U^\mu U_\nu+\delta^{\mu}_\nu\mathrm{p}\ , 
\end{equation}
and then use (\ref{P}) to write down the 4-velocity to second order in the metric perturbations:
\begin{equation}\label{4vel}
U^0=a^{-1}e^{-\Psi}(1+v_i\omega^i+v^2/2)\ , \ \ \quad U^i=a^{-1}e^{\Phi}v^i\ .
\end{equation}
This last method is easiest to work with, and by using it for the baryons and the dark matter we obtain:
\begin{eqnarray}
&&T^i_{\ 0} = - e^{\Psi + \Phi} (\rho+\mathrm{p}) v^i\ , \nonumber\\
&&T^i_{\ j} = \delta^i_j \mathrm{p} + (\rho+\mathrm{p}) v^i v^j\ .
\end{eqnarray}
The above  $T^i{}_{ 0}$ is correct also for the photons upon replacement of $\vec v$ with $\vec v_\gamma$. The result is different from the one in \cite{bartolo}.
From (\ref{eq:Tij}), we recover the result of \cite{bartolo} for  $T_\gamma^{\, i}{}_{\; j}$:
\be
T_{\gamma}^{\, i}{}_{\; j} = \bar{\rho}_{\gamma} \left( \Pi^i_{\gamma j} + \frac{1}{3}
\delta^i_j \left( 1 + \Delta_{00}^{(1)} + \frac{\Delta_{00}^{(2)}}{2} \right)
\right)\ .
\end{eqnarray}

\subsection{The second order Einstein equation}

We now give the Einstein tensor, $G^\mu_{\ \nu}$, to second order in the Poisson gauge:
\begin{eqnarray}
G^0_{\ 0} &=& - \frac{e^{- 2 \Psi}}{a^2} \left[ 3\mathcal{H}^2 - 6\mathcal{H}
\dot{\Phi} + 3 \dot{\Phi}^2 - e^{2 \Phi + 2 \Psi} \left( \Phi_{, i} \Phi_{, i}
- 2 \Phi_{, k k} \right) \right]= \nonumber\\&&= \kappa^2 T^0_{\ 0}\ ,\\
G^i_{\ 0} &=& 2 \frac{e^{2 \Phi}}{a^2} \left[ \dot{\Phi}_{, i} + (\mathcal{H}-
\dot{\Phi}) \Psi_{, i} \right] - \frac{1}{2 a^2} \omega^i_{, k k} + \left(
4\mathcal{H}^2 - 2 \frac{\ddot{a}}{a} \right) \frac{\omega^i}{a^2} = \nonumber\\&&= \kappa^2
T^i_{\ 0}\ ,\\
G^i_{\ i} &=& \frac{3 e^{- 2 \Psi}}{a^2} \left[ \mathcal{H}^2 - 2 \frac{\ddot{a}}{a}
- 2 \dot{\Phi} \dot{\Psi} - 3 \dot{\Phi}^2 + 2\mathcal{H} \left( \dot{\Psi} +
2 \dot{\Phi} \right) + 2 \ddot{\Phi} \right]\nonumber \\
&&+ \frac{e^{2 \Phi}}{a^2} \left[ 2
\Psi_{, k} \Psi_{, k} - 2 \Psi_{, i} \Phi_{, i} + \Phi_{, i} \Phi_{, i} + 2
\left( P - 3 N \right) \right] \\ \nonumber
&&= \kappa^2 T^i_{\ i}\ ,\\
P^{ij}_{kl}G^k_{\ l}&=&\frac{e^{2\Phi}}{a^2}P^{ij}_{kl}\left(\Phi_{,kl}-\Psi_{,kl}+\Phi_{,k}\Phi_{,l}-\Psi_{,k}\Psi_{,l}-\Psi_{,k}\Phi_{,l}-\Phi_{,k}\Psi_{,l}\right)\nonumber\\
&&-a^{-2}\left[\frac{\dot{\omega}_{i,j}+\dot{\omega}_{j,i}}{2}+\mathcal{H}(\omega_{i,j}+\omega_{j,i})\right]+
\frac{1}{2a^2}[\ddot{\chi}_{ij}-\nabla^2\chi_{ij}+2\mathcal{H}\dot{\chi}_{ij}]\nonumber\\
&&=\kappa^2 P^{ij}_{kl}T^k_{\ l}\ ,
\end{eqnarray}
with the projection operator defined as $P^{ij}_{kl}\equiv[\delta_{ik}\delta_{jl}-\delta_{ij}\delta_{kl}/3]$, and $\kappa^2\equiv 8\pi G_N$. The equations above are equivalent to the ones in App. A of \cite{bartolo2} with $\Phi_{\mathrm{here}}=\Psi_{\mathrm{there}}$ and $\Psi_{\mathrm{here}}=\Phi_{\mathrm{there}}$. The transverse part of the last equation above gives the gravitational waves generated by the scalar mode \cite{baumann}.

The  remaining longitudinal part of $G^i_{\ j}=\kappa T^i_{\ j}$ is given by
\be
\Phi-\Psi=\mathcal{Q}\ ,
\ee
where ${\mathcal{Q}}$ is defined through the following chain of definitions:
\be
 \nabla^2 \mathcal{Q}= 3 N-P\ ,
\ee
with
\be
P=P^i_{\ i}\ , \quad {\rm and}\quad  \nabla^2 N=P^i_{\ j,ij}\ , 
\ee
where finally:
\begin{eqnarray}
P^i_{\ j}=\Phi_{,i}\Psi_{,j}+\frac{1}{2}(\Psi_{,i}\Psi_{,j}-\Phi_{,i}\Phi_{,j})+\Phi(\Psi_{,ij}-\Phi_{,ij})+4\pi G_Na^2T^i_{\ j}\ .
\end{eqnarray}
Note that our expression for $P^i_{\ j}$ is equivalent to the one in \cite{bartolo2}.

\section{Summary\label{section:summary}}

The purpose of this paper was to provide a detailed analysis of the perturbations to the recombination history of the universe and give the full corrected second order Boltzmann and Einstein equations for the photons, baryons and CDM for a flat FRW background. Those are necessary for the correct treatment of CMB secondaries coming from recombination.

We analyzed the different timescales in recombination and argued that a perturbed version of the recombination equation for the Peebles 3-level atom is adequate for obtaining the evolution of the perturbations to the electron density. We derived rigorously the perturbations to the escape probability by angle averaging the photon distribution function, and showed that it is sufficient to treat the escaping of Ly$\alpha$ photons as only due to the local velocity divergence. 

We showed that for modes longer than the timescale of recombination (corresponding to $k\lesssim0.1\,$Mpc$^{-1}$), it is a good approximation to consider $n_e^{(0+1)}$ as determined by the local value of $n_b^{(0+1)}$ (or equivalently of the temperature), with the same relationship as in an unperturbed universe: $\delta n_e\simeq \delta n_b\, \dot n_e/\dot n_b$. This is due to the fact that the timescale of recombination is much faster than the considered modes. For the same reason,  this also implies that the amplitude of the electron density perturbations is enhanced by a factor of roughly five compared to the baryon density perturbations. This enhancement is physical since it persists for modes well inside the horizon, down to scales comparable with the photon diffusion scale (corresponding to $k\lesssim0.2\,$Mpc$^{-1}$). The enhancement is most prominent around the peak of the photon visibility function and therefore has the potential to have an impact on the CMB bispectrum, as we will investigate in a companion paper \cite{inprep}. 

Solving for the density of free electrons in the universe, as well as writing the full set of second order Boltmann and Einstein equations requires to be careful with the definition of coordinates and collision terms. For this reason, we reviewed the formulation of kinetic theory in curved spacetime and applied it to derive the Boltzmann and Einstein equations at second order in the perturbed FRW universe. We compared our results with former pre-existing literature. We find that  the former derivations are not completely correct, and we give the necessary corrections.

Solving the full set of second order equations is a very hard task, which however is motivated by the current experimental and theoretical status. In a companion paper \cite{inprep}, we solve approximately these second order equations restricting ourself to modes well inside the horizon and to second order sources that are proportional to $\delta_e$, and we compute the induced bispectrum. There we find that the non-Guassian signal is equivalent to about $f_{\rm NL}\sim-5$, which is potentially detectable by future experiments such as Planck.

\section*{\large Note Added}
While this paper and its companion \cite{inprep} were slowly written down,  preprint \cite{Pitrou:2008hy} appeared very recently. Our results about the full set of second order equations in sec.~\ref{section:secondorder} partially overlap with those of \cite{Pitrou:2008hy}.

\section*{Acknowledgments}

The work of LS is supported in part  by the National  Science Foundation under Grants No. PHY-0503584.

\begin{appendix}

\section{Covariant Form of the Boltzmann Equation}\label{section:A}
\subsection{Preliminaries}
In this Appendix we intorduce some useful quantities in covariant kinetic theory following closely the work of Ehlers \cite{ehlers}, to which we will often refer for details. We work in generic coordinates, and we define the affine parameter of each particle trajectory so that
\be
 \frac{d x^\mu}{d\lambda}=P^\mu\ ,
\ee
where $P^\mu$ is the physical momentum. In practice $\lambda$ of one particle  is defined so that $d \lambda$ coincides with the time interval of a local inertial observer divided by the energy observed in the same frame: $d\lambda=dt_{\rm inertial}/E_{\rm inertial}$.
Let us first introduce the diff. invariant measure of the one-particle phase space. The measure in momentum phase space is given by:
\begin{eqnarray}\label{pi}
\pi  \equiv g_{\mathrm{deg}} 2\sqrt{-g}d^3P^i\int dP^0 \delta(P^\mu P_\mu-m^2)\theta(P^0)=g_{\mathrm{deg}}\frac{\sqrt{-g}}{|P_0|}d^3P^i\nonumber\\
  =g_{\mathrm{deg}}\frac{2d^3P_i}{\sqrt{-g}}\int dP_0 \delta(P^\mu P_\mu-m^2)\theta(P_0) =g_{\mathrm{deg}}\frac{d^3P_i}{|P^0|\sqrt{-g}}\ ,
\end{eqnarray}
where $g_{\mathrm{deg}}$ are the internal degrees of freedom (degeneracy) of the particle species for which $\pi$ is calculated. The relationship between the momenta in generic coordinates and the momenta in the local inertial frame is given by:
\begin{equation}\label{tetrad_P}
P^\mu=e^\mu_{\ \nu'}p^{\nu'}\ .
\end{equation}
Here, $p^0=-p_0=E$ is the instantaneous energy of the particle with respect to the local observer. The tetrad satisfies the set of equations $e^\mu_{\ \alpha'}e^\nu_{\ \beta'}g_{\mu\nu}=\eta_{\alpha'\beta'}$. For example, in the particular case in which the generic coordinates are the ones of Newtonian gauge, they are given at second order in eq.~(\ref{tetrad2}).
From  (\ref{tetrad_P}), we can rewrite $dP^i=[e^i_{\ 0}\frac{p_{j}}{E}+e^i_{\ j}]dp^{j}$ using $EdE=p_jdp^j$. Plugging this expression for $dP^i$ in (\ref{pi}) and using $P_0=g_{0\mu}e^\mu_{\ \nu'}p^{\nu '}=\eta_{\nu'\sigma'}\left(e^{-1}\right)^{\sigma'}_{\ 0}p^{\nu'}$, it is straightforward to show that 
\begin{equation}\label{pi-local}
\pi=g_{\mathrm{deg}}\frac{d^{3}p}{E}\ .
\end{equation}
We find that, once expressed in terms of momenta in the inertial frame, the momentum-space measure reduces to the standard Lorentz invariant one.
The one-particle phase space needs also the measure for the spacetime volume. This is given by the standard 4-form $\varpi\equiv \sqrt{-g}d^4x$, where $g\equiv\det g_{\mu\nu}$. The one-particle phase space is 7-dimensional, and its measure is given by $\Omega=\varpi\, \pi$. Notice that both the spacetime volume measure and the momentum-space measure are separately diff. invariant.

\subsection{Invariant distribution function and the Boltzmann equation}

The Liouville operator $L$ is a vector acting on the 7-dimensional space $(x^\mu,P_i)$. It is given by
\be
L=\frac{ d}{d\lambda}=P^\mu\frac{\d}{\d x^\mu}+\frac{d P^i}{d\lambda}\frac{\d}{\d P^i}\ .
\ee
$L$ clearly describes the free streaming of a particle in the 7-dimensional phase space.
If we contract $L$ with $\Omega$ we obtain a 6-form: 
$$\omega= L\cdot \Omega\ .$$ 
Up to a constant factor, which will be fixed shortly, this is the unique 6-form which assigns a nonzero volume to any hypersurface not tangent to the vector flow $L$. By construction, $\omega$ is invariant with respect to the phase flow, i.e. $\int_\Sigma\omega=\int_{\Sigma'}\omega$, where $\Sigma$ and $\Sigma'$ are two cross-sections of a tube, defined by the Liouville flow. Doing the contraction in the expression for $\omega$ one obtains
\begin{equation}\label{omega_restr}
\omega=P^\mu\sigma_\mu\pi\ ,
\end{equation}
where $\sigma_\alpha$ is the covariant surface element~\footnote{Given a set of coordinates $(u^1,u^2,u^3)$ parametrizing a 3-surface, the 3-surface measure can be written as $\sigma_\alpha=(1/3!)\sqrt{-g}\epsilon_{\alpha\beta\gamma\delta}(\partial x^\beta/\partial u^1) (\partial x^\gamma/\partial u^2)(\partial x^\delta/\partial u^3)d^3u$ with $\epsilon_{\alpha\beta\gamma\delta}=\epsilon_{[\alpha\beta\gamma\delta]}$, and $\epsilon_{0123}=1$ \cite{Weinberg}.}. For our physical interests,  we restrict $\omega$ to be defined on surfaces $\Sigma$ of the kind $\Sigma= G\times K$, where $G$ is a spacelike hypersurface, and $K$ is the mass shell hyperboloid in momentum space. Going to a local inertial frame instantaneously at rest with a particle with 4-momentum $P^\mu$, we see that $\omega$ reduces to $d^3 xd^3 p$. Thus $\omega$ is the standard 6-dimensional phase-space volume measure, a fact that fixes the remaining constant that made the definition of $\omega$ not unique.

If we call $N[\Sigma]$ the average number of 7-dimensional trajectories that cross a certain surface $\Sigma$, we can define the phase-space one-particle distribution $f(x^i,P_i;\eta)$ through the relationship
\be
N[\Sigma]=\int_{\Sigma} f\, \omega\ .
\ee
By varying $\Sigma$ and making it arbitrarily small, and by using the uniqueness of $\omega$, one can show that $f$ is a uniquely defined scalar function \cite{ehlers}.

The average number of interactions in a phase space region $D$, $N[\partial D]$, is given by 
\begin{equation}\label{coll_old}
N[\partial D]=\int_{\d D} f\, \omega=\int_{D} d(f\omega)=\int_{D} L[f]\Omega\ .
\end{equation}
In the last passage we have used that $d(f\omega)=L[f]\Omega$ (see \cite{ehlers}). Thus, the phase-space density of collisions is given by $L[f]=df/d\lambda$. With this interpretation of $L[f]$, we can finally write the Boltzmann equation in covariant form: 
\be\label{boltzmannequation}
L[f]=C[f]\ ,
\ee
where $C[f]$ is the collision term giving the phase-space density of collisions, which we will discuss in the next subsection. Before moving on, let us notice that in the above equation one can use $L[f]=df/d\lambda=P^0df/d\eta$ to obtain
\be\label{boltz1}
P^0\frac{df}{d\eta}=C[f]\ ,
\ee
which is the form of the Boltzmann equation which we use in most of the paper.

Notice that we can rewrite (\ref{coll_old}) as:
\be
N[\partial D]=\int_G \sigma_\mu \int_K \pi\, P^\mu f\ ,
\ee
which allows us to read off the expression for the particle current as 
\be\label{current}
{\cal{N}}^\mu=\int_K \pi P^\mu f \ .
\ee
We notice that its divergence is given by~\cite{ehlers}:
\be\label{coll}
{\cal{N}}^\mu_{\ ;\mu}=\int_K L[f]\pi\ .
\ee

%Using Stokes' theorem, we can rewrite (\ref{coll_old}) as 
%\begin{equation}
%N[\partial D]=\int_{\partial D} f\omega\ .
%\end{equation}
%Consider $D$ to be a segment of a flux tube bounded on the top and bottom by ``caps'' given by the union of two spacelike hypersurfaces, $G\times K$. Since $\omega$ vanishes along the tube boundary, then only the caps contribute to the above integral. Thus, we can finally write
%\begin{equation}\label{NpD}
%N[\partial D]=\int_{G} \sigma_\mu \int_K P^\mu f \pi \ ,
%\end{equation}
%where we used (\ref{omega_restr}). The above integral gives the difference between the number of particle trajectories intersecting the upper and lower caps of the flux tube, which exactly equals the number of interactions in $D$ as expected. 

%Using Gauss' theorem for a given spacetime volume, $S$, from (\ref{coll_old}) and (\ref{coll}) with $D=S\times K$ we can write
%\be
%N[\partial D]=\int\limits_{S} N^\alpha_{\ ; \alpha}\varpi=\int\limits_{\partial S} N^\alpha \sigma_\alpha\ ,
%\ee
%and comparing with (\ref{NpD}), we can read off the expression for the particle 4-current:
%\begin{equation}\label{current}
%N^\mu=\int_K f P^\mu \pi
%\end{equation}

\subsection{The collision term}

Having introduced the invariant distribution function, $f$, and the Boltzmann equation (\ref{boltz1}), we can procede with the discussion of the collision term, $C[f]$, which is a scalar by construction. Let us consider interactions of the kind $iK_1,jK_2\cdots\to rK_3\dots$, where $\{i,j,\cdots,r,s,\dots\}$ label the species of the particles, and $K_a$ labels the momentum regions for each particle. The collision term for $f_i$ is then given by \cite{ehlers}
\begin{eqnarray}\label{C}
C[f_i]&=&\int_{K_2}\pi_j\cdots\int_{K_3}\pi_r\cdots\nonumber\left[f_i(x^k,P_1;\eta )f_j(x^k,P_2;\eta )\cdots(1\pm f_r(x^k,P_3;\eta ))\cdots\right]\nonumber\\
&&\times W(g_{\mu\nu};iP_1,jP_2,\cdots;rP_3,\cdots)\ ,
\end{eqnarray}
where $P_n$ is the 4-momentum of the $n^{\mathrm{th}}$ particle participating in the collision. Notice that in the equation above there is obviously no integration over $\pi_i$. Since the phase space measures are diff. invariant, so it is $W(g_{\mu\nu};\{ P\})$. We can write it in a local inertial frame $W(g_{\mu\nu};\{P\})=W(\eta_{\mu\nu};\{p\})$, where the transformation between $p$ and $P$ is given by the tetrad in eq.~(\ref{tetrad_P}). In a local inertial frame $W(\eta_{\mu\nu};{p})$ is expressible using the usual scattering operator $S$:
\begin{eqnarray}\label{W}
W(\eta_{\mu\nu};ip_1,jp_2,\cdots;rp_3,\cdots)&\equiv&(2\pi)^{3m-4}\delta^{(4)}(p_1+p_2+\cdots-p_3-\cdots)\nonumber\\
&&\times\left|\left<rp_3,\cdots\right|M\left|ip_1,jp_2,\cdots\right>\right|^2\, ,\nonumber\\
\left<rp_3,\cdots\right|S-1\left|ip_1,jp_2,\cdots\right>&\equiv&\delta^{(4)}(p_1+p_2+\cdots-p_3-\cdots)\nonumber\\
&&\times\left<rp_3,\cdots\right|M\left|ip_1,jp_2,\cdots\right>\ , 
\end{eqnarray}
where $m$ is the total number of particles in the interaction, and $p_i$ is the momentum of the $i^{\mathrm{th}}$ particle.

\subsection{Nice coordinates}
Notice that the collision term is a scalar local quantity in spacetime, and therefore, if we write it in the frame where the momenta are the inertial ones, then it does not contain any metric perturbation. This is a nice simplification that we wish to exploit. In order to do this,  we can define a new distribution function $F(x^\mu,p^i)$ as:
\be
F(x^\mu,p^i)=f(x^\mu,P^i(p^j,x^\nu))\ .
\ee
The new distribution function is defined everywhere in spacetime, and in each point it is a function of the local inertial frame momenta defined by eq.~(\ref{tetrad_P}). We call this `mixed' set of phase space coordinates `nice coordinates'~\footnote{This is what is usually done, in a less formal way, when studying the Boltzmann equation at first order \cite{Dodelson:2003ft}
. The generalization to second order can therefore be a source of confusion. Here we are explaining how to perform correctly this change of variables directly at general non-linear level.}. In fact, in these coordinates the collision term $C[f]$ becomes just the standard Minkowski one, with no metric fluctuations. In particular, the momenta measure just becomes
\be
\pi_{i}=g_{\rm deg,\, i}\frac{d^3p_i}{E_i} \ ,
\ee
and the metric which contracts the momenta becomes $\eta_{\mu\nu}$.
Coming now to the Liouville term
\be
\frac{d\, f(x^\mu,P^i(p^j,x^\nu)}{d\lambda}\ ,
\ee
it is easy to write it in terms of $F(x^\mu,p^i)$ by applying the chain rule. Since the Liouville term is a total derivative with respect to $\lambda$, this simply amounts in doing
\be
\frac{d\, f(x^\mu,P^i)}{d\lambda}=\frac{d\, F(x^\mu, p^i)}{d\lambda}\ .
\ee
The Boltzmann equation in nice coordinates becomes:
\begin{eqnarray}
\frac{\partial F}{\partial \eta }+\frac{\partial F}{\partial x^i}\frac{dx^i}{d\eta }+\frac{\partial F}{\partial p^i}\frac{dp^i}{d\eta }=\frac{1}{P^0}C[F]\ ,
\end{eqnarray}
which is the version we use in the paper.
All the dependence on the metric is now just inside $dx^i/d\eta$, $dp^i/d\eta$ and $P^0$.

\section{Derivation of equation (\ref{delta_K_P})}\label{section:B}

\subsection{Einstein coefficients}
We are now going to derive the expression for the escape probability at first order in the fluctuations. In order to do this, it is useful to first derive an expression for the Einstein coefficients using the notation we introduced in the previous sections. For absorption of a photon by an atom in a transition from the $i^\mathrm{th}$ to the $j^\mathrm{th}$ level, we have the following spacetime density of interactions
\begin{eqnarray}\label{Bij_gen}
\left.{\cal{N}}^{\mu}_{;\mu}\right\vert_{\rm Abs.}=\int \pi_i\pi_j\pi_\gamma f_\gamma f_i W(i,\gamma;j)\ ,
\end{eqnarray}
where $f_i$ denotes the one-particle distribution function of the atoms in the $i^\mathrm{th}$ level, and $f_\gamma$ is the one-particle distribution function for the photons.
Going to the inertial frame instantaneously at rest with respect to the the atoms ($\vec{ v}_b=0$), i.e. to the laboratory frame, the result for the spacetime density of absorptions is given by the following term of the standard radiative transfer equation 
\begin{eqnarray}\label{Bij}
\left.{\cal{N}}^{\mu}_{;\mu}\right\vert_{\rm Abs.}=B_{ij}n_i\int I(\nu,\hat n) \phi_{ij}(\nu,\hat n)d\nu d\Omega/4\pi=g_{\gamma}\frac{B_{ij}}{4\pi}n_i\int  \nu F_\gamma \phi_{ij}(\nu, \hat n)d^3 p\ ,
\end{eqnarray}
where $g_{\gamma}=2$ is the degeneracy of the photons, $B_{ij}$ is the Einstein coefficient for absorption, $\nu$ and $p$ are respectively the energy and momentum of the photons measured by an inertial observer for whom the bulk velocity of the atoms is zero, $\phi_{ij}(\nu,\hat n)$ is the laboratory absorption line profile, satisfying $\int \phi_{ij}(\nu,\hat n)d\nu d\Omega/4\pi=1$, and $n_i$ is the proper number density of atoms in the $i^\mathrm{th}$ level. In (\ref{Bij}) we used $I(\nu,\hat n)=g_\gamma \nu^3F_\gamma$, where $I$ is the light intensity. 

Equating  (\ref{Bij_gen}) and (\ref{Bij}), we obtain an expression for $B_{ij}$ in terms of a collision term:
\begin{equation}\label{Bijfinal}
\frac{\nu^2_{ij}}{4\pi}\phi_{ij}(\nu,\hat n)B_{ij}n_i=\left.\int\pi_i\pi_j f_i W(i,\gamma;j)\right|_{\vec{ v}_b=0 \mathrm{\ frame}}=g_ig_j\left.\int\frac{d^3 p_i}{E_i}\frac{d^3 p_j}{E_j} F_i W(i,\gamma;j)\right|_{\vec{ v}_b=0 \mathrm{\ frame}}\ ,
\end{equation}
where $g_i$ and $g_j$ are the degeneracies of the $i^{\mathrm{th}}$ and $j^{\mathrm{th}}$ levels, respectively.
The integral over $\pi_\gamma$ can be dropped because this equation is valid for any $f_\gamma$.

The derivation of the $A_{ji}$ and $B_{ji}$ Einstein coefficients proceeds in an analogous way. The spacetime number density of downward transitions is given by:
\begin{eqnarray}
&&\left.{\cal{N}}^{\mu}_{;\mu}\right\vert_{\rm Em.}=B_{ji}n_j\int I(\nu,\hat n) \phi_{ji}(\nu,\hat n)d\nu d\Omega/4\pi+A_{ji}n_j\int\phi_{ji}(\nu,\hat n)d\nu d\Omega/4\pi =\\ \nonumber
&&=g_{\gamma}\frac{B_{ji}}{4\pi}n_j\int  \nu F_\gamma \phi_{ji}(\nu, \hat n)d^3 p+\frac{A_{ji}n_j}{4\pi\nu_{ij}^2}\int\phi_{ji}d^3 p=
\frac{B_{ji}}{4\pi}n_j\nu_{ij}^2\int\phi_{ji}\left(F_\gamma+\frac{A_{ji}}{B_{ji}g_\gamma\nu_{ij}^3}\right)\pi_\gamma\ .
\end{eqnarray}
But it can be also read off from the Boltzmann equation:
\begin{eqnarray}
\left.{\cal{N}}^{\mu}_{;\mu}\right\vert_{\rm Em.}=\left.\int\pi_i\pi_j\pi_\gamma F_j(1+F_\gamma)W(j;i,\gamma)\right|_{\vec{ v}_b=0 \mathrm{\ frame}}\ .
\end{eqnarray}
Equating these two expressions we end up with an analogous expression for $B_{ji}$ and $A_{ji}$:
\begin{eqnarray}
&&\frac{\nu^2_{ij}}{4\pi}\phi_{ji}(\nu,\hat n)B_{ji}n_j=\left.\int\pi_i\pi_j f_j W(j;i,\gamma)\right|_{\vec{ v}_b=0 \mathrm{\ frame}}
=g_ig_j\left.\int\frac{d^3 p_i}{E_i}\frac{d^3 p_j}{E_j} F_j W(j;i,\gamma)\right|_{\vec{ v}_b=0 \mathrm{\ frame}}\label{Bji}\ ,\nonumber\\
&&\left.\int\pi_i\pi_j f_j(1+f_\gamma) W(j;i,\gamma)\right|_{\vec{ v}_b=0 \mathrm{\ frame}}
=\frac{\nu^2_{ij}}{4\pi}\phi_{ji}(\nu,\hat n)B_{ji}n_j\left(\frac{A_{ji}}{B_{ji}g_\gamma\nu_{ij}^3}+f_\gamma\right)\label{1pfg}\ .
\end{eqnarray}
The last expression gives just the standard relation $A_{ji}= g_\gamma \nu^3 B_{ji}$. We can also see that the laboratory profiles for stimulated and spontaneous emission are identical, given by $\phi_{ji}(\nu,\hat n)$~\footnote{ This holds in the semi-classical approximation to first order in the intensity \cite{CHO}.}.

\subsection{Perturbations to the escape probability}

The escape probability is the only non-local interaction during recombination which enters in the simplified Peebles treatment, so it needs a special discussion. In the homogeneous universe calculation in Section \ref{section:lyhom} we derived the escape probability from the Boltzmann equation which we in turn obtained from the detailed balance of the number of photons. In this section we will instead  start directly with the Boltzmann equation.

The Boltzmann equation (\ref{Boltzmann}) and the expression for the collision term (\ref{C}) written for the photons give 
\begin{eqnarray}
\frac{df_\gamma}{d\lambda}=\int \pi_i\pi_j\left[f_j (1+f_\gamma)W(j;i,\gamma)-f_if_\gamma W(i,\gamma;j)\right]\ .
\end{eqnarray}
Using (\ref{Bijfinal}) and (\ref{1pfg}) we can rewrite the above equation in the inertial frame at rest with the baryons to first order as
\begin{eqnarray}\label{photonBoltz_curved}
\left.\frac{d F_\gamma}{d \lambda}\right|_{\vec{ v}_b=0\  \mathrm{frame}}=\frac{1}{8\pi\nu}[A_{ji}n_j-2\nu^3(B_{ij}n_i-B_{ji}n_j)F_\gamma]\phi(\nu)\ ,
\end{eqnarray}
where $\lambda$ is an affine parameter, and $\vec{v}_b$ is the baryon bulk velocity. Note that in a homogeneous universe the above equation reduces to (\ref{photonBoltz}).

As in the homogeneous case, we can reparametrize time by using the frequency $\nu(t)$ of a photon as measured by inertial observers at rest with the baryons. Since we are working in a perturbed FRW, we need to specify a direction of propagation $\hat{n}$ for the photons. Thus, all quantities in eq.~(\ref{photonBoltz_curved}) acquire a dependence on $\hat{n}$ further than the dependence on $\nu$. We can integrate it to obtain the analogous of eq.~(\ref{inttrans}):
\begin{eqnarray}\label{inttrans2}
&&F_\gamma(\nu,\hat n)=F_\gamma(\nu_B,\hat n)e^{-\tau(\nu,,\hat n)}+e^{-\tau(\nu,\hat n)}\int_{\nu_B}^{\nu}e^{\tau(\nu',\hat n)}\frac{S(\nu',\hat n)}{2\nu'^3}\frac{d\tau(\nu',\hat n)}{d\nu'}d\nu'\ , 
\end{eqnarray}
which just comes from the integration on each separate direction of eq.~(\ref{Fdiff}).
In order to compute the evolution of the number density of electrons (see for example eq.~(\ref{Bij})), we just need an expression for the angle averaged escape probability. Since we are working to first order, and all quantities are angle independent at zeroth order, in the Sobolev approximation, we find that the angle averaging factorizes as follows
\begin{eqnarray}
&&\left<\bar F_\gamma\right>_\Omega=P\left< F_{bb}\right>_\Omega+(1-P)\frac{\left<S\right>_\Omega}{2\nu_{ij}^3}\ , \nonumber\\ \label{eq:weight}
&& P\equiv  \int^\infty_0d\nu\langle\phi(\nu,\hat{n})\rangle_\Omega \exp[-\langle\tau(\nu,\hat{n})\rangle_\Omega]= \langle\gamma\rangle_\Omega\left(1-e^{-1/\langle\gamma\rangle_\Omega}\right)\ ,
\end{eqnarray}
where
\begin{eqnarray}
&&\gamma(\nu,\hat{n})\equiv -\frac{d\nu}{d\lambda}\frac{4\pi}{\nu^2(B_{ij}n_i-B_{ji}n_j)}\ .\label{gammapert}
\end{eqnarray}
Here angular brackets denote angular averaging, and a bar denotes frequency averaging over the line profile (i.e. $\int d\nu\phi(\nu)$). This is the analogous of eq.~(\ref{eq:simpleF}) that we found in the unperturbed case. Notice also that $\gamma$ reduces to the homogeneous result when $d\nu/d\lambda=-H\nu^2$ as in the unperturbed Universe. In the Sobolev approximation the angular dependence in $\gamma$ enters only through the factor $d\nu/d\lambda$, and therefore  we need to find its angle average to first order. Let us rewrite it as
\begin{eqnarray}\label{temp999}
\frac{d\nu}{d\lambda}=
%\frac{d\lambda(\mathrm{fixed }\ \vec{ x} \ \mathrm{frame})}{d\lambda(\vec{  v}_b=0 \ \hbox{frame})}P^0\frac{d\nu}{d\eta}=
P^0\frac{d\nu}{d\eta}\ .
\end{eqnarray}
Here $\nu$ is measured in the rest frame of the atoms (we are avoiding the subscript $_{\vec v_b=0}$ for clarity), while $P^0$ is the momentum measured by a comoving observer sitting at fixed $\vec{ x}$. If we introduce the momentum measured in a LIFIRCO $p$, the transformation between $p$ and $\nu$ is just a Doppler shift, which to first order is given by $p=\nu(1+\vec{  v}_b\cdot \hat{ n})$.  Combining (\ref{P}) with (\ref{temp999}) we obtain
\begin{equation}
a e^\Psi P^0\frac{d\nu}{d\eta}=p\frac{d\nu}{d\eta}=\nu(1+\vec{v}_b\cdot \hat{ n})\frac{d(p(1-\vec{  v}_b\cdot \hat{n}))}{d\eta}=\nu(1+\vec{  v}_b\cdot\hat{n})\left(\frac{dp}{d\eta}(1-\vec{  v}_b\cdot\hat{n})-\frac{d\vec{  v}_b}{d\eta}\cdot\hat{n}\nu\right)\ .
\end{equation}
Combining this with the geodesic equation in the Newtonian gauge (see e.g. \cite{maed} or (\ref{dpdt})), to first order we obtain
\begin{eqnarray}\label{dnudt1}
ae^\Psi P^0\frac{d\nu}{d\eta}=-\mathcal{H}\nu^2\left[1-\frac{\dot \Phi}{\mathcal{H}}+\hat n^i\frac{\Psi_{,i}}{\mathcal{H}}+ v_{b,i}\hat n^i+\hat n_i\cdot\frac{d v_b^i}{d\eta}\frac{1}{\mathcal{H}}\right]\ .
\end{eqnarray}
All first order vector quantities can be written as the gradient of a potential, i.e. $\vec{A}=\vec\nabla A$. Here we are interested in distances $\xi$ small compared to the typical gradients of the first order quantities, and thus, when angle averaging, we can Taylor expand around the origin:$$\vec A(\vec x)=\vec{A}(\vec{x}=0)+\xi \hat n^i\d_i \vec A(\vec x=0)\ .$$ Notice that we are ignoring the time dependence of $\vec A$. If included, this would give similar effects to the ones we will find as due to the space dependence, which as we will see shortly, give negligible contribution. 
When angle averaging $n^i A_i(\xi \hat n)$, therefore, we obtain $$\langle\xi n^i n^j \d_i\d_j A(\vec{x}=0)\rangle_{\Omega}=\frac{\xi \d_i A^i(\vec x=0)}{3}\ .$$
%All first order vector quantities are proportional to $\vec{ A}\propto \hat k e^{i\vec{ k}\cdot\vec{x}}$. Expanding $\vec{A}(\vec{\xi})$ with $\vec{\xi}=\xi \hat{n}$ to first order we get in Fourier space $\vec{A}(\vec{\xi})=\vec{A}(\vec{x}=0)+i\xi(\hat{\vec{ n}}\cdot \vec{ k})\vec{A}(\vec{ x}=0)$. Thus, it can be easily checked that the angle average of $n^iA_{i}$ is given by $\left<n^iA_{i}\right>= i \xi\vec{ k}\cdot \vec{A}(\vec{x}=0)/3$.  
Thus, after angle averaging (\ref{dnudt1}), we find
\begin{eqnarray}\label{angleave}
\left<P^0\frac{d\nu}{d\eta}\right>_\Omega=-\frac{\mathcal{H}\nu^2}{a}\left[1-\Psi-\frac{\dot \Phi}{\mathcal{H}}+\frac{\xi}{3}\frac{\nabla^2\Psi}{\mathcal{H}}+\frac{\xi}{3}\theta_b+\frac{\xi}{3\mathcal{H}}\dot \theta_b+\frac{1}{3\mathcal{H}}\theta_b \right]\ ,
\end{eqnarray}
where we denoted $\theta_b\equiv \d_j v_b^j$ and used $d\vec{  v}_b/d\eta=\dot{\vec{  v}}_b+\vec{ v}_{b\, ,i}dx^i/d\eta=\dot{\vec{  v}}_b+n^i\vec{ v}_{b\, ,i}$ (where we applied that to zeroth order $dx^i/d\eta=\hat n^i$, since $\vec v_b$ is already first order). In (\ref{angleave}), because of the weighting with the line profile  in (\ref{eq:weight}), $\xi$ is order $L_S$. Since $\xi$ multiplies first order quantities, the above equation holds since to zeroth order we have $d\xi/d\eta=1$. Comparing the above equation with eq.~(\ref{delta_K_P}) we can see that the perturbation to the escape probability has acquired contributions other than the local velocity divergence, $U^\mu_{;\mu}/3=H$. Combining (\ref{angleave}) with the momentum conservation equation for the baryons in Newtonian gauge \cite{maed}, we obtain
\be
\left<P^0\frac{d\nu}{d\eta}\right>_\Omega=-\frac{\nu^2}{3}\left[ U^\mu_{;\mu}+a^{-1}\xi\left(-c_s^2\nabla^2\delta_b+\frac{4}{3}\frac{\rho_\gamma}{\rho_b}an_e\sigma_T(\theta_\gamma-\theta_b)\right)\right]\ .
\ee
Thus, the additional terms in (\ref{angleave}) correspond to the change of the momentum of the baryon fluid due to pressure forces and Thomson scattering in a physical time interval equal to $a\xi$. The terms multiplied by $\xi$ are nonlocal effects due to the photons sampling the Sobolev patch.

The Sobolev approximation holds only when the perturbation scale $k^{-1}$ is much larger than the Sobolev scale. Comparing numerically the terms in (\ref{angleave}) and using $\xi\sim L_S\sim 5$ kpc we identify the non-negligible first order terms. We find that within the Sobolev timescale, the change in the baryon momentum due to pressure forces and Thomson scattering is negligible. Therefore, we can write
\begin{eqnarray}\label{dnudlambda}
\left<\frac{d\nu}{d\lambda}\right>_\Omega=-\frac{\mathcal{H}\nu^2}{a}\left(1-\Psi-\frac{\dot \Phi}{\mathcal{H}}+\frac{\theta_b}{3\mathcal{H}}\right)\ .
\end{eqnarray}

Combining (\ref{dnudlambda}) and (\ref{gammapert}) we find that the angle-averaged Sobolev parameter is given by
\begin{eqnarray}
\langle\gamma\rangle_\Omega=\frac{4 \pi U^\mu_{\ ;\mu}}{3(B_{ij}n_i-B_{ji}n_j)}\ .
\end{eqnarray}
This confirms the former intuitive derivation of the perturbation $\delta H$ to the escape probability as given in (\ref{delta_K_P}).
% where we used the first order number conservation for the baryons (\ref{num_cons_1}).

\section{Errata to the second order Compton collision term}\label{section:C}
Although the second order expansion of the Compton collision term is present in many references, we find that often there are typos (actually we did not find one reference which is fully correct). For this reason, we prefer to summarize the corrections we found to what is reported in  \cite{dodjub} and in \cite{bartolo}. Let us start with Ref.~  \cite{dodjub}.
\begin{itemize}
\item In the third line of (2.9) in \cite{dodjub}, there is an extra prime in the second factor of the last term, i.e. it should read $... \ -f^{(0)}(p')[1+f^{(0)}(p)] \ ...$.

\item In the second line of (2.13) in \cite{dodjub}, the last term in the curly brackets should be multiplied by $T_e$, so it should read $\left\{...\ -T_e[f^{(0)}(p')-f^{(0)}(p)]\partial\delta(p-p')/\partial p'\right\}$

\item In (2.15) in \cite{dodjub}, the first factor should not be present, i.e. the equation should read $c^{(2)}_{\Delta v}(\vec{p},\vec{p}')=[f^{(1)}(\vec{p}')-f^{(1)}(\vec{p})] \ ...$
\end{itemize}

Using the Compton collision term from \cite{dodjub}, with the above corrections, we checked the final expression for $C(\vec{p})$ given in \cite{bartolo} by their eq.~(4.42). There is  one correction that needs to be made there: 
\begin{itemize}
\item  If one looks at eq.~(4.12) of \cite{bartolo},  some of the above terms have been corrected, but still  in our opinion there are some typos remaining. Still, in eq.~(4.42) only one typo is remaining. This is the last term, the Kompaneets one, which should be multiplied by $m_e/p^2$, giving the standard $\frac{1}{m_e p^2}\frac{\partial}{\partial p}\ ...$.
\end{itemize}

For computing the temperature fluctuations at second order, we find it useful to write the Collision term in a generic frame:
%\begin{scriptsize}
\be\label{compton2}
&&\frac{C^{(2)}_{e\gamma}(F_\gamma)}{2 n_e\sigma_T p_\gamma}=\frac{1}{2}F_{00}^{(2)}(p)-\frac{1}{4}\sum_m \sqrt{\frac{4 \pi}{5^3}}F_{2m}^{(2)}Y_{2m}(\hat n)-\frac{1}{2}F^{(2)}(\vec p) +\\ \nonumber 
&&\delta_e^{(1)}\left[F_{00}^{(1)}-F^{(1)}-\frac{1}{2}\sum_m \sqrt{\frac{4\pi}{5^3}}F_{2m}^{(1)}Y_{2m}(\hat n)-p\frac{\partial F^{(0)}}{\partial p}\vec v\cdot\hat n\right]
-\frac{1}{2}p\frac{\partial F^{(0)}}{\partial p}\vec v^{(2)}\cdot \hat n\\\nonumber 
&&+\vec v\cdot \hat n\left[F^{(1)}-F^{(1)}_{00}-p\frac{\partial F^{(1)}_{00}}{\partial p}-\frac{1}{2}\sqrt{\frac{4\pi}{5^3}}\sum_m\left(F_{2m}^{(1)}-p\frac{\partial F^{(1)}_{2m}}{\partial p}\right)Y_{2m}(\hat n)+i\sqrt{\frac{\pi}{3}}\sum_m F_{1m}^{(1)}Y_{1m}(\hat n)\right]\\ \nonumber  &&
+
2\pi v\sqrt{\frac{2}{15}}\sum_{m,M}\left(
\begin{array}{ccc}
1 & 1 & 2 \\
m & M & -m-M
\end{array}
\right)F_{2,m+M}^{(1)}Y_{1m}(\hat n)Y_{1M}(\hat v)(-1)^{m+M}\\ \nonumber 
&&-i\sqrt{\frac{\pi}{3^3}}v \sum_m Y_{1m}(\hat v)\left(5 F_{1m}^{(1)}+2 p\frac{\partial F^{(1)}_{1m}}{\partial p}\right)\\ \nonumber  &&
+\frac{2^{3/2}\pi v}{15}\sum_{l=1,3}(-i)^l(-1)^m\left(\delta_{l,1}-\frac{3}{14^{1/2}}\delta_{l,3}\right)\left(\begin{array}{ccc}
1 & 2 & l \\
-m & -M & m+M
\end{array}
\right)\times \\ \nonumber
&&\times Y_{1m}(\hat v)Y_{2M}(\hat n)\left(4F_{lm}^{(1)}+p\frac{\partial F^{(1)}_{lm}}{\partial p}\right)\\
\nonumber 
 &&
+(\vec v\cdot \hat n)^2\left[p\frac{\partial F^{(0)}}{\partial p}+\frac{11}{20}p^2\frac{\partial^2 F^{(0)}}{\partial p^2}\right]+
v^2\left[p\frac{\partial F^{(0)}}{\partial p}+\frac{3}{20}p^2\frac{\partial^2 F^{(0)}}{\partial p^2}\right]\\ \nonumber
&&+\frac{1}{m_ep^2}\frac{\partial}{\partial p}\left[p^4\left(T_e\frac{\partial F^{(0)}}{\partial p}+F^{(0)}(1+F^{(0)})\right)\right]\ .
\ee
%\end{scriptsize}
Here, following the notation of BMR1, we have expanded the distribution function as
\be
F=F^{(0)}+F^{(1)}+\frac{1}{2}F^{(2)}\ ,
\ee
where $f^{(1)}$ is the first order perturbation and $f^{(2)}$ is the second order one. The expansion in sperical harmonics of a quantity $f(\hat n)$ is given by:
\be
f_{lm}=(-i)^{-l}\sqrt{\frac{2l+1}{4\pi}}\int d\Omega f(\hat n)Y^\star_{lm}(\hat n)\ .
\ee

\end{appendix}

\end{document}